\RequirePackage{amsmath}
\newcommand\newblock{\hskip .11em\@plus.33em\@minus.07em}
\documentclass[12pt,a4paper,final]{iopart}

\usepackage{graphicx}
\usepackage{physics}
\usepackage{siunitx}
\usepackage[dvipsnames]{xcolor}
\usepackage{soul}

\usepackage[square,numbers]{natbib}
\usepackage{hyperref}
\usepackage{doi}

\begin{document}
\title[Optical cycling of AlF molecules]{Optical cycling of AlF molecules}
\author{S Hofs\"{a}ss$^1$, M Doppelbauer$^1$, S C Wright$^1$, S Kray$^1$, B G Sartakov$^2$, J~P\'{e}rez-R\'{i}os$^1$, G Meijer$^1$, S Truppe$^1$}

\address{$^1$Fritz-Haber-Institut der Max-Planck-Gesellschaft, Faradayweg 4-6, 14195 Berlin, Germany}
\address{$^2$Prokhorov General Physics Institute, Russian Academy of Sciences, Vavilovstreet 38, 119991 Moscow, Russia}
\ead{truppe@fhi-berlin.mpg.de}

\begin{abstract}
Aluminium monofluoride (AlF) is a promising candidate for laser cooling and trapping at high densities. We show efficient production of AlF in a bright, pulsed cryogenic buffer gas beam, and demonstrate rapid optical cycling on the Q rotational lines of the $A^1\Pi \leftrightarrow X^1\Sigma^+ $ transition. We measure the brightness of the molecular beam to be $>10^{12}$ molecules per steradian per pulse in a single rotational state and present a new method to determine its velocity distribution in a single shot. The photon scattering rate of the optical cycling scheme is measured using three different methods, and is compared to theoretical predictions of the optical Bloch equations and a simplified rate equation model. Despite the large number of Zeeman sublevels (up to 216 for the Q(4) transition) involved, a high scattering rate of at least $17(2)\times 10^6$ s$^{-1}$ can be sustained using a single, fixed-frequency laser without the need to modulate the polarisation. \textcolor{black}{We deflect the molecular beam using the radiation pressure force and measure an acceleration of $8.7(1.5)\times 10^5$ m/s$^2$}. Losses from the optical cycle due to vibrational branching to $X^1\Sigma^+, v''=1$ are addressed efficiently with a single repump laser. Further, we investigate two other loss channels, parity mixing by stray electric fields and photo-ionisation. The upper bounds for these effects are sufficiently low to allow loading into a magneto optical trap. 
\end{abstract}

\vspace{2pc}

\section{Introduction}
Laser cooling and trapping of atomic gases has transformed fundamental physics research and has enabled the invention of precise instruments, such as atomic clocks, magnetometers, gravimeters and accelerometers. Ultracold molecules are expected to have a similar profound impact on science and technology \cite{Carr2009, Safronova2018}. Their rich energy level structure and long-range dipolar interactions offer many new opportunities, but also make them more challenging to cool. Over the last decade, there has been substantial progress in developing techniques to cool molecules or to form them by associating ultracold, laser-cooled atoms.
Stark deceleration \cite{Bethlem1999} combined with adiabatic cooling \cite{Cheng2016} as well as opto-electrical Sisyphus cooling \cite{Prehn2016} can be used as very general tools to increase the phase-space density of molecules. Zeeman deceleration \cite{Vanhaecke2007a} combined with deep magnetic traps offer the potential for sympathetic and evaporative cooling \cite{Segev2019}. This method is as general as Stark deceleration and gives access to a complementary class of molecules. The efficient association of laser-cooled atoms to ultracold molecules has recently produced the first quantum degenerate gas of polar bi-alkali molecules \cite{DeMarco2019}, followed by the observation of sympathetic and evaporative cooling \cite{Son2020, valtolina2020dipolar}. This paved the way to use field-induced collisional resonances between molecules to avoid inelastic collisions \cite{Matsuda2020}. To extend the variety of molecular species that can be cooled to ultralow temperatures, atomic laser cooling techniques have been adapted to diatomic \cite{Shuman2010} and polyatomic \cite{Mitra2020} molecules, demonstrating magneto-optical trapping and sub-Doppler cooling for a variety of molecular species \cite{Barry2014, Truppe2017, Anderegg2018, Lim2018, McNally2020, Ding2020}. The density of the molecules captured and cooled in a magneto-optical trap (MOT) is still at least four orders of magnitude lower compared to atomic MOTs. This has not impeded the loading of optical \cite{Anderegg2018, Anderegg2019} and magnetic traps \cite{Williams2018, McCarron2018}, to study molecular collisions \cite{Cheuk2020} and long rotational coherence times \cite{Caldwell2020}. However, the low density in molecular MOTs severely limits the subsequent use of laser-cooled molecules to study strongly interacting, many-body quantum systems. 

The low density is a result of the low number of molecules delivered to the MOT at velocities low enough to be captured. So far, only the molecular radicals SrF, CaF and YO with a $X^2\Sigma^+$ ground state have been laser-cooled in a MOT. Their high reactivity limits the obtainable density in the molecular source. In addition, to prevent optical pumping into dark rotational levels, the P(1) rotational line of the $A^2\Pi_{1/2}\leftrightarrow X^2\Sigma^+$ electronic transition must be used for laser cooling. The large excess of ground-state levels involved in the optical cycle results in a photon-scattering rate that is typically 10 times lower compared to alkali atoms. This leads to a long slowing distance and the characteristically low capture velocity of molecular MOTs of around $10$ m/s. The unpaired electron in the electronic ground state results in a large magnetic $g$-factor which severely complicates the use of a Zeeman slower \cite{Petzold2018}. Furthermore, the lifetime of the molecules in the MOT is limited by optical pumping into dark ro-vibrational levels, which prevents accumulating molecules over long time-periods. In addition, the molecular beam used to load the MOT is typically pulsed, so that only a single molecular pulse is loaded into the MOT. 

Recently, we have identified the AlF molecule as an excellent candidate to overcome these limitations \cite{Truppe2019}. It is inherently more stable than the radicals commonly used for molecular MOTs, produces an intense molecular beam and allows implementing a fast optical cycling scheme that provides a strong spontaneous scattering force which is essential for a high-density MOT. Here, we characterise the molecular beam source and devise a simple and fast method to measure the velocity distribution of the molecules. The short lifetime of the first electronically excited state of AlF (1.9 ns), the highly diagonal Franck-Condon matrix and the specific energy level structure allow us to implement a simple, highly closed and fast optical cycling scheme for all Q lines of the $A^1\Pi\leftrightarrow X^1\Sigma^+$ transition near 227.5 nm. We show that, despite the complex hyperfine structure a high photon scattering rate ($> 17 \times 10^6$ s$^{-1}$) can be obtained. The radiation pressure force exerted onto the molecules is measured by deflecting the molecular beam and compared to the photon scattering rate obtained from fluorescence measurements and from optical pumping to $X^1\Sigma^+, v''=1$. Numerical simulations of the optical cycling process are in good agreement with the experimental results. A rate-equation model describes the optical cycle well for low laser intensities, but fails to predict the scattering rate accurately at high intensities. After scattering about 200 photons, the molecules are lost from the main cooling cycle to $v''=1$. These molecules are then recovered with a high efficiency by using a repump laser, closing the optical cycle and leaving only a small leak to $X^1\Sigma^+, v''=2$. Finally, we investigate losses arising electric field-induced parity mixing in the excited state and from two-photon ionisation. Neither of these will inhibit loading a MOT.

\section{Energy level structure and optical cycling scheme}
\subsection{{\label{sec:elecBranching}}Electronic levels}
The molecules that have been laser-cooled so far (apart from YO) consist of an alkaline earth and alkaline earth-like atom bonded to a fluorine, hydrogen or hydroxyl ligand to form a $^2\Sigma^+$ ground state. These radicals have an unpaired electron which is localized at the metal atom and laser excitation promotes it without changing the bond length of the molecule. This leads to the very diagonal Franck-Condon matrix, which is typical for these molecules, and as a result the number of vibrational repump lasers required is two or three. 

\begin{figure}[tb]
\centering
\includegraphics[width=0.7\linewidth]{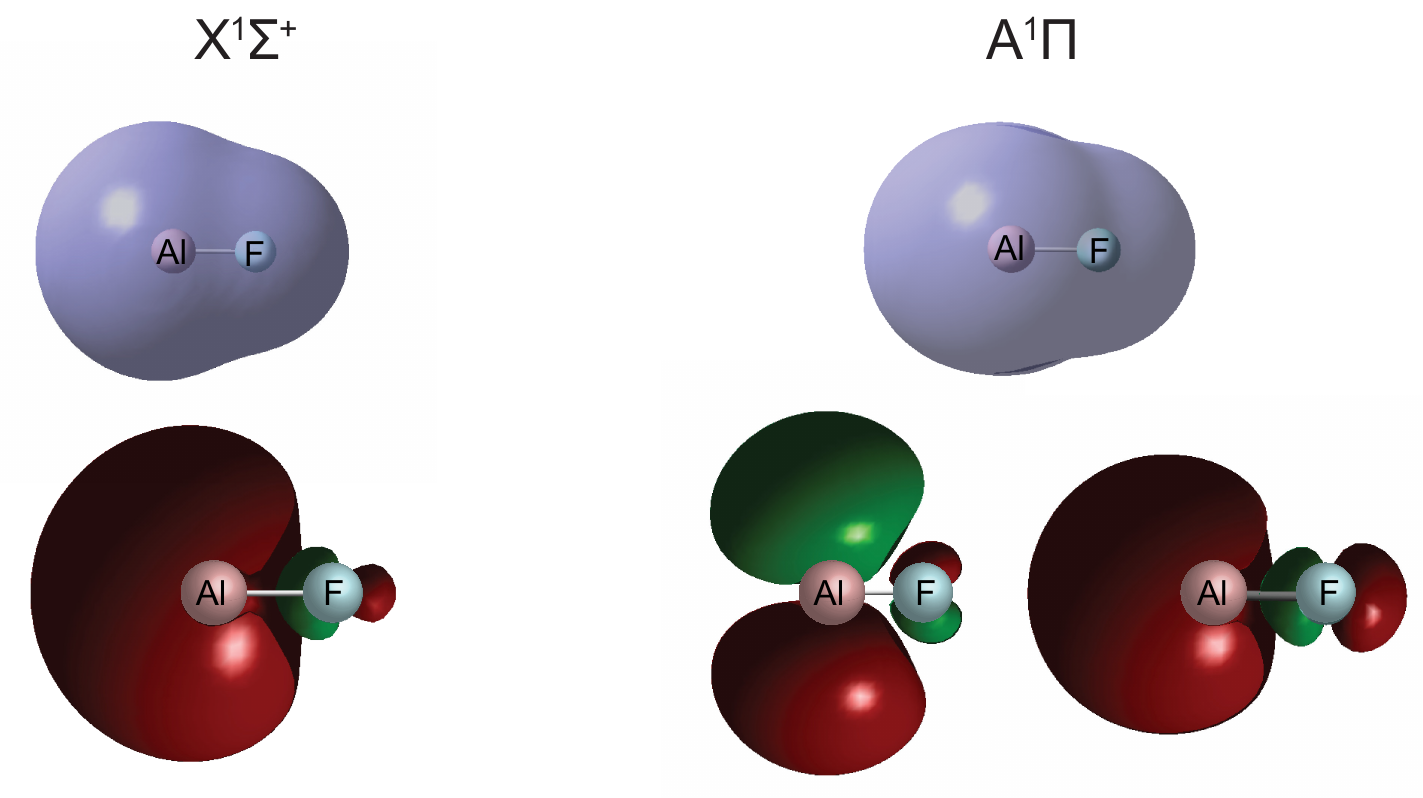}
\caption{\label{fig:orbitals} Calculated electron density (top) and molecular orbitals of AlF (bottom). The electron density is very similar in the two electronic states and the electronic excitation resembles an atomic transition between an s and p orbital. The molecular bond changes only minimally which limits vibrational branching and therefore the number of repump lasers required for laser cooling.}
\end{figure}

The situation is slightly different for AlF, which combines a group III element with fluorine to form a $X^1\Sigma^+$ ground state. The AlF bond is mostly ionic\footnote{ \textcolor{black}{The definition of the ionic character of a molecular bond is not unique ~\cite{vanArkel1956, Pauling1986, Mori2002, Meek2005, Hou2015, Pototschnig2016}. Depending on the definition that is used, the bond in AlF is characterized as mostly ionic or at the border between polar covalent and ionic.}}, with the 3p electron of Al being largely transferred to the fluorine atom \cite{Dearden1991}. \textcolor{black}{Laser excitation promotes one electron from the $7\sigma^2$ molecular orbital, which has primarily Al 3s character \cite{So1974}, into a $3\pi$ orbital which has primarily Al 3p character}. This results in a $A^1\Pi$ or a $a^3\Pi$ state, depending on the orientation of the spin of the electron. We have performed electronic structure calculations to better understand why the Franck-Condon matrix is diagonal, despite the lack of an unpaired electron. Figure \ref{fig:orbitals} shows the calculated electron density and molecular orbitals of the electronic states relevant to this study. The electron density (top of each panel) has been calculated within the multi-reference-configuration-interaction (MRCI) method available in MOLPRO 2019.2 \cite{MOLPRO}.\footnote{The MRCI calculations are fed with the natural orbitals resulting from a Multi-Configuration Self-Consistent Field (MCSCF) calculation with a complete active space (CAS) consisting of nine orbitals with $A_1$ symmetry, three with $B_1$ symmetry, and three with $B_2$ symmetry associated with the point group $C_{2v}$.} The bottom part of each panel shows the HOMO of the $X^1\Sigma^+$ electronic state, and the HOMO and LUMO for the $A^1\Pi$ electronic state, respectively. The calculations are carried out by employing the AVQZ \cite{BasisSet} basis set for each atom. 

The electron density in the $X^1\Sigma^+$ and $A^1\Pi$ state is very similar and the electronic excitation resembles that of an atomic transition between and s and p orbital \cite{Ivanov2019}. It involves only a minimal change in the molecular bond which results in a diagonal Franck-Condon matrix and allows optical cycling with only a few vibrational repump lasers. 

Spontaneous decay on the spin-forbidden $A^1\Pi \rightarrow a^3\Pi$ transition is weakly allowed, due to spin-orbit mixing. This causes a small leak from the optical cycle below the 10$^{-6}$ level which we have analysed in detail previously \cite{Truppe2019, Doppelbauer2020}.

\subsection{{\label{sec:vibBranching}}Vibrational levels}

\begin{figure}[tb]
\centering
\includegraphics[]{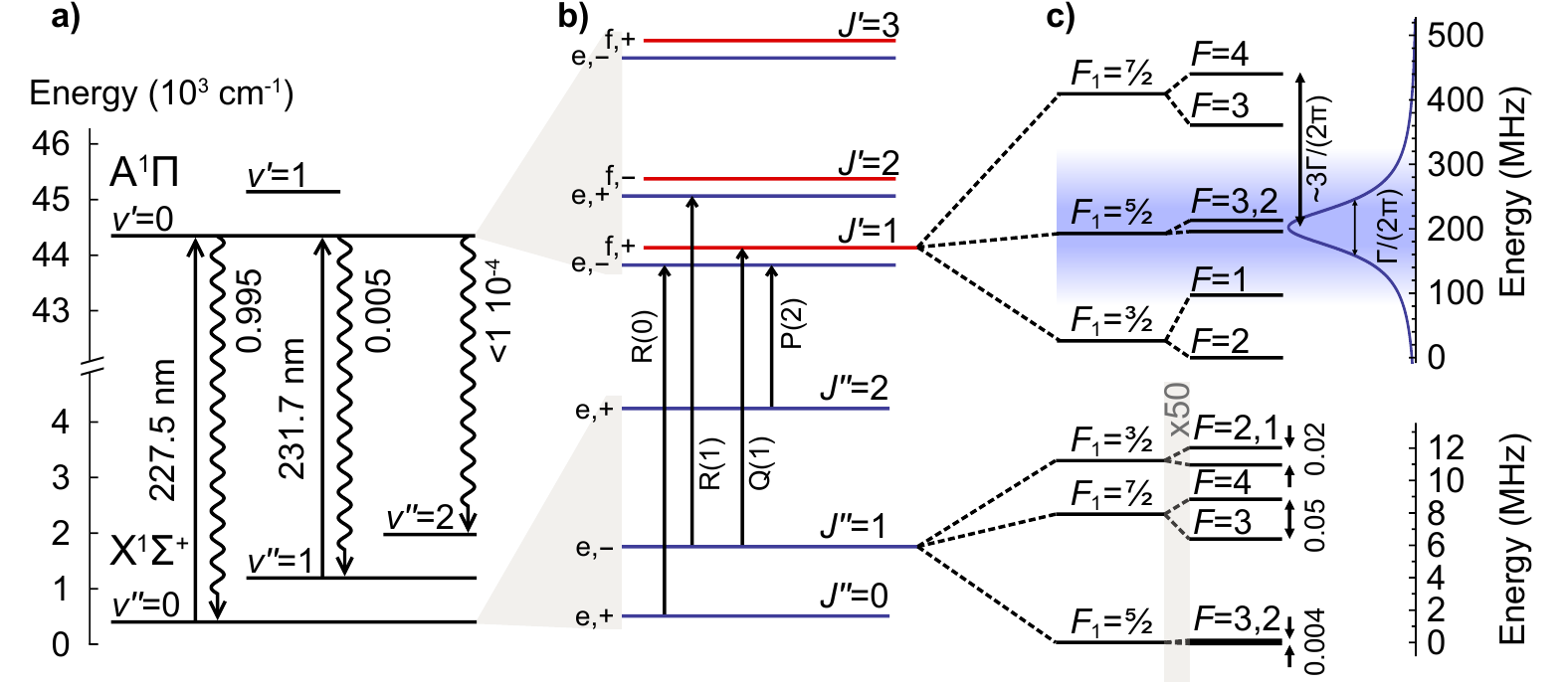}
%multifigure is in nextcloud cycling paper/figures/energyscheme_rev4
\caption{\label{fig:levelscheme} Energy level diagram of the states relevant for laser cooling of AlF. a) Electronic and vibrational states of the pump ($v'=0 \leftarrow v''=0$) and repump ($v'=0 \leftarrow v''=1$) transitions (solid arrows). Wavy arrows indicate spontaneous emission and the calculated branching ratios are given.
b) $\Lambda$-doubling in the excited state results in two opposite parity states per rotational level. R$(J'')$ and P$(J'')$ lines reach e levels while Q$(J'')$ lines reach f levels and are rotationally closed.
c) The aluminium nuclear spin splits $J=1$ levels into three $F_1$ components, each of which is split again by the fluorine nuclear spin into a total of six $F$ levels. The hyperfine structure in the ground state is small compared to the natural linewidth $\Gamma/(2\pi)=84$~MHz and completely unresolved. \textcolor{black}{The splitting of the $F_1$ components is magnified by a factor of 50.}}
\end{figure}

Figure \ref{fig:levelscheme} a) shows the relevant vibronic energy levels for the optical cycle of AlF, together with the transition wavelengths and calculated branching ratios. The spontaneous decay rate from $A^1\Pi,v'=0\rightarrow X^1\Sigma^+,v''=0$ is proportional to $A_{00}$, the Einstein $A$ coefficient of the $0-0$ band. Molecules that can be laser-cooled typically have $A_{00}\gg A_{01}$ \cite{DiRosa2004}. Without any repump lasers to address the vibrational branching to $v''>0$ the average number of photons that can be scattered is determined by $r=A_{00}/\sum^{\infty}_{v''=0}A_{0v''}=\tau_0 A_{00}$, where $r$ is the probability for a molecule to decay back to $v''=0$ and $\tau_0=1.9$ ns is the lifetime of $A^1\Pi,v'=0$ level. This is often approximated by the Franck-Condon factor which neglects the variation of the transition dipole moment with the internuclear distance. 

\textcolor{black}{The probability distribution for a molecule to scatter exactly $n$ photons, when the number is limited to $N$, is $p(n)=(1-r)r^{n-1}$ for $n<N$, $p(n)=r^{n-1}$ for $n=N$, and $p(n)=0$ for $n>N$. For a molecule that interacts with a laser for a time $t_i$ the maximum number of photons scattered is $N=R t_i$, where $R$ is the photon scattering rate. When an ensemble of molecules interacts with a laser beam, the fraction of molecules remaining in the initial state is given by $P(N)=r^N$ \cite{DiRosa2004}. The average number of photons scattered is
\begin{align}
    \langle n_{\textrm{ph}}\rangle=\sum^{N}_{n=0}n p(n)=\frac{1-r^N}{1-r}=\frac{1-P(N)}{1-r}.
    \label{eq:nphot}
\end{align}
}
The limiting value for $N\rightarrow\infty$ is given by $\langle n_\textrm{ph}^\infty\rangle=1/(1-r)$, and the standard deviation of \textcolor{black}{the probability distribution} is equal to the mean value. This means that there is a wide variation of the number of photons scattered by \textcolor{black}{individual} molecule\textcolor{black}{s} \cite{DiRosa2004, McNally2020}. We can approximate $r$ by the previously calculated ratio of Einstein coefficients $r=1/(1+A_{01}/A_{00})=0.9953$ \cite{Truppe2019} and arrive at a \textcolor{black}{maximum} mean number of scattered photons of $\langle n_\textrm{ph}^\infty\rangle=213$.\footnote{Depending on the basis set and potential that is used to calculate the branching ratio we estimate its uncertainty to be about 15\%.} The population in $v''=0$ decreases with $N$, according to $P_{v''=0}(N)=r^N$, which for $r\approx 1$ is well approximated by an exponential decay $P_{v''=0}(N)\simeq e^{-N(1-r)}$. In the limit of $\Gamma t_{i} \gg 1$ and $R\ll\Gamma$, with $\Gamma=1/\tau_0$ being the spontaneous decay rate, 
\begin{align}
    n_{\textrm{ph}}(t_i)=\frac{1-e^{-R (1-r)t_i}}{1-r}. 
    \label{eq:nphotPop}
\end{align}
The limit of $\langle n_\textrm{ph}^\infty\rangle$ can only be reached if $R(1-r)t_i\gg 1$ and the fluorescence saturates \cite{Tarbutt2009}. 

\subsection{\label{sec:rotBranching}Rotational levels}
For convenience, we describe both, the ground and excited state, using Hund's case (a). The angular momenta $\mathbf{R}$, $\mathbf{L}$ and $\mathbf{J}=\mathbf{R}+\mathbf{L}$ describe the end-over-end rotation of the nuclei, the total orbital angular momentum of the electrons and the total angular momentum excluding hyperfine interactions, respectively.\footnote{\textcolor{black}{There is no spin angular momentum in the ground and excited state and therefore $\mathbf{J}\equiv\mathbf{N}=\mathbf{R}+\mathbf{L}$. For simplicity we use $J$ as a quantum number for both the ground and excited state.}} The relevant rotational structure is shown in figure \ref{fig:levelscheme} b). To prevent rotational branching it is common to use the strict parity and angular momentum selection rules of electric dipole transitions \cite{Stuhl2008}. In AlF, each Q line of the $A^1\Pi\leftrightarrow X^1\Sigma^+$ transition is rotationally closed. The number of photons a molecule can scatter is in this case limited by losses to higher vibrational levels. 

The rotationally open P and R transitions can be used as ``standard candles'' to calibrate the laser-induced fluorescence of the cycling transition. The average number of photons scattered by a molecule is limited by optical pumping to dark rotational states that are not coupled to the laser light. Analogous to the vibrational branching discussed above, the rotational branching ratios determine the number of photons a molecules scatters. Contrary to the vibrational case, the branching ratio for the rotational transitions $r=r_{\Delta J}(J)$ depends on the transition type and $J''$ and is determined by the Hönl-London factor $S_J^{\Delta J}(J'')$, which for a $^1\Pi-{}^1\Sigma^+$ transition is given by
\begin{alignat}{2}
    S^{P}(J'')&=\frac{1}{2} J''- \frac{1}{2} \qquad &\text{for }& J''\geq2 \\
    S^{Q}(J'')&= J''+ \frac{1}{2} \qquad &\text{for }& J''\geq1\\
    S^{R}(J'')&= \frac{1}{2} J''+ 1 \qquad &\text{for }& J''\geq0,
\end{alignat}
\textcolor{black}{where} $\sum_{\Delta J}{S^{\Delta J}}(J'')=2J''+1$ is the total degeneracy \cite{kovacs1969rotational}. The branching ratios can be written as  
\begin{alignat}{1}
    r_P(J'')&=\frac{S^{P}(J'')}{S^{P}(J'')+S^{R}(J''-2)}\\
    r_Q(J'')&=1\\
    r_R(J'')&=\frac{S^{R}(J'')}{S^{R}(J'')+S^{P}(J''+2)}, 
\end{alignat}
with a limiting value for the number of photons scattered by a molecule $\langle n^{\infty}_{\textrm{ph,}P} \rangle=2-1/J''$ and $\langle n^{\infty}_{\textrm{ph},R}\rangle=2+1/(J''+1)$ for P and R lines, respectively. There is an additional vibrational loss of about 1/213 per cycle, which lowers the expected number of photons for each line by about 1\%. When including the hyperfine structure the limiting values must be corrected to account for additional hyperfine dark states which can lower the expected number of scattered photons by about 10\%.

\subsection{{\label{sec:theory}}Hyperfine levels}
\begin{figure}[tb]
\centering
\includegraphics[]{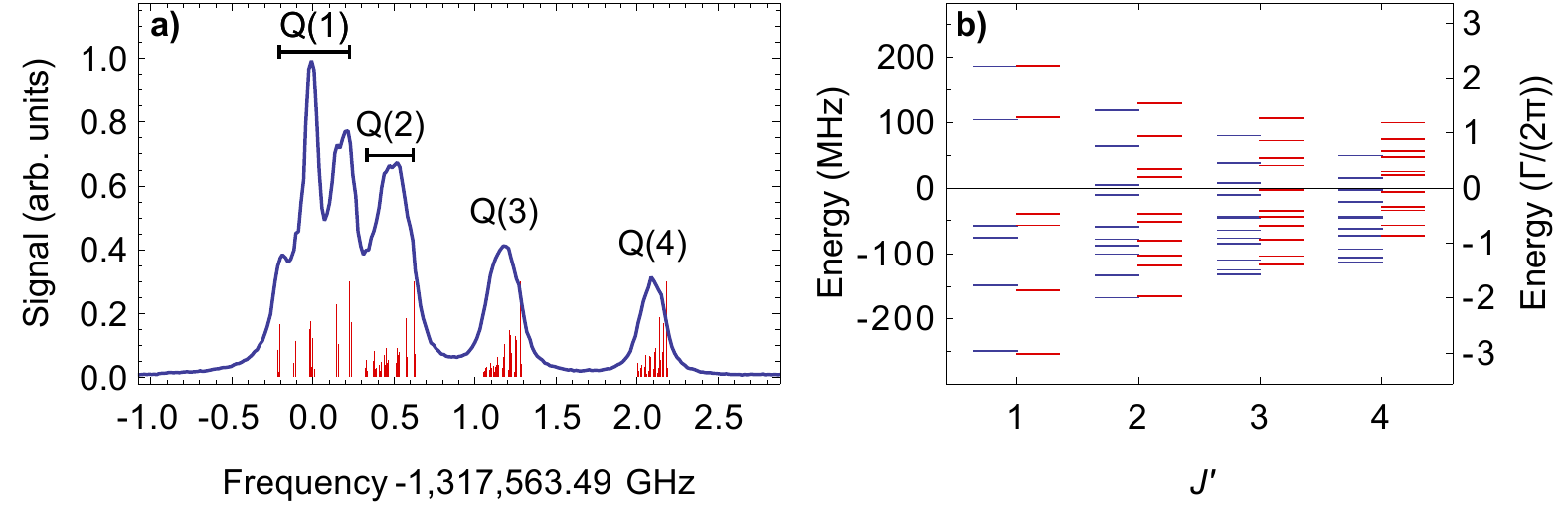}
\caption{\label{fig:qbranchhfs} a) Laser-induced fluorescence spectrum of the Q-branch of the $A^1\Pi, v'=0\leftarrow X^1\Sigma^+,v''=0$ band. The sticks below the spectrum represent the individual hyperfine transitions and are calculated using the spectroscopic parameters from \cite{Truppe2019}. Panel b) shows the hyperfine splittings of the $A^1\Pi, v'=0, J'$ levels with respect to their gravity center. The $e$ levels are blue and $f$ levels are in red.}
\end{figure}

\begin{figure}[tb]
\centering
\includegraphics[]{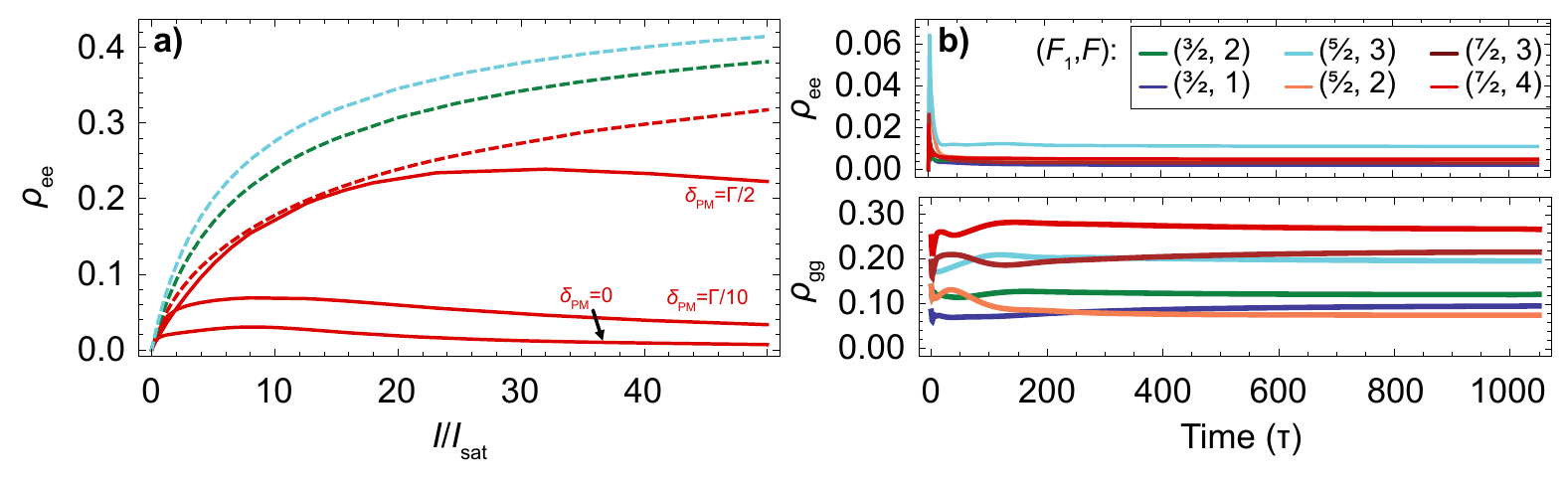}
\caption{{\label{fig:obe}} Comparison of the optical cycling rate derived from solving a rate-equation model (dashed) and the optical Bloch equations (solid) in the absence of any losses. a) Excited state population $\rho_{ee}$ as a function of the saturation parameter $s=I/I_{\textrm{sat}}$, where $I_{\textrm{sat}}=\pi h c \Gamma/(3\lambda^3)=0.93$ W/cm$^2$. The solution to the rate-equation model for the Q(1), Q(2) and Q(3) are shown in red, green and light blue, respectively. The simulations using optical Bloch equations (solid red) for the Q(1) line are shown for three different polarisation modulation rates $\delta_{\textrm{PM}}$. For all the simulations, the laser frequency is set to the centre of each rotational line. The optical cycling rates on the Q(1) line obtained from the two theoretical models agree well at low intensities, but there is a significant difference at high intensities. b) Time-evolution of the populations in the various hyperfine levels in the $X^1\Sigma^+, v''=0, J''=1$ state for $\delta_{\textrm{PM}}=0$ and $s=8$.}
\end{figure}

Both the aluminium nucleus and the fluorine nucleus have an unpaired proton. Therefore, both carry a nuclear spin, given by $I_{\textrm{Al}}=5/2$ and $I_{\textrm{F}}=1/2$ and combines to the total angular momentum $\mathbf{F}=\mathbf{F}_1+\mathbf{I}_\textrm{F}$, where $\mathbf{F}_1=\mathbf{J}+\mathbf{I}_{\textrm{Al}}$. The hyperfine interaction splits the $J=0, 1$ and 2 levels into two, six, and ten hyperfine levels, respectively, and into 12 $F$-levels for $J\geq 3$. 

Figure \ref{fig:levelscheme} c) presents the relevant energy level structure for $J'=1$ and $J''=1$. In the electronic ground state, the hyperfine structure is dominated by the interaction of the electric quadrupole moment of the Al nucleus with the local electric field gradient. As a result, the intermediate angular momentum $\mathbf{F}_1=\mathbf{J}+\mathbf{I}_{\textrm{Al}}$ is a good quantum number.  The total span of the hyperfine structure in the $X^1\Sigma^+$ state lies well within the natural linewidth $\Gamma/(2\pi)=84$ MHz of the $A-X$ transition, where $\Gamma=\sum_{v''} A_{0v''}=1/\tau_0$ is the spontaneous decay rate. \textcolor{black}{This means that radiofrequency sidebands or additional lasers to cover the spin-rotation or hyperfine splittings in the ground state are not required.}

In the excited state, each $J'$-level comprises of a closely-spaced pair of opposite parity levels, a $\Lambda$-doublet, with $\Lambda$ being the projection of $\mathbf{L}$ onto the internuclear axis. Each $\Lambda$-doublet component is split by the hyperfine interaction. The interaction arising from the Al and F nuclei is comparable and $F_1$ is not a good quantum number. The total span of the hyperfine intervals in the lowest $J'$ level is about 50 times larger ($\approx 6 \Gamma/(2\pi)$) than in the ground state and reduces to $\approx 2 \Gamma/(2\pi)$ for $J'=4$. The decreasing span of the hyperfine structure in the $A$~state can be seen in the Q-branch spectrum of the $A^1\Pi, v'=0\leftarrow X^1\Sigma^+,v''=0$ band provided in figure \ref{fig:qbranchhfs} a). The hyperfine structure in the $A^1\Pi, v'=0$ level is partially resolved for the Q(1) line only. Panel b) gives a more detailed representation of the hyperfine levels of the $A^1\Pi, v'=0$ level for $J'=1$ to $J'=4$ with respect to their gravity centre by using the parameters from our previous spectroscopic study of AlF \cite{Truppe2019}. 

According to \cite{Berkeland2002}, a $J'=J''\leftrightarrow J''$ transition, such as the cycling Q lines discussed here, has one dark state for any choice of polarisation in zero magnetic field. Assuming linearly polarised light and the quantisation axis along the polarisation axis, the molecules are pumped into the dark $\ket{J'',M_{J''}=0}$ state after scattering only a few photons. When a molecule is pumped into such a dark state it ceases to fluoresce because it does not couple to the excitation light anymore. This is detrimental for Doppler cooling.\footnote{Dark states are not always detrimental. Robust, velocity selective dark states can be very useful to cool molecules to very low temperatures below the recoil limit \cite{Caldwell2019}.}

In general, a dark state can either be an angular momentum eigenstate or a coherent superposition of these eigenstates. The former type is stationary, i.e. it does not evolve in time, and molecules accumulate and remain there indefinitely. The latter type of dark state can be non-stationary, i.e. it precesses between a dark and a bright state. We determine the dark state composition by using equation 23 in \cite{Fitch2021} and find that for linear polarisation they are superpositions of states with different values of $F_1$. The smallest $F_1$ splitting is $\sim \Gamma/30$ (see Figure \ref{fig:levelscheme} c)) and for a fixed linear polarisation it is this separation that limits the scattering rate (see below). 

A common method to increase the scattering rate is to lift the degeneracy of the ground states by inducing a Zeeman splitting of the order of $\Gamma$ \cite{Berkeland2002}. To achieve this for AlF a magnetic field of a few T is needed, due to the small magnetic $g$-factor in the $^1\Sigma^+$ ground state. A second method is to rapidly switch the polarisation of the light. Each polarisation is associated with a set of dark states, and switching the polarisation recovers a high scattering rate, provided the number of ground states is less than three times the number of excited states.  For the Q-lines of AlF, the number of ground states is equal to the number of excited states, and all are accessible with a single laser frequency. As a result, polarisation modulation can be applied to increase the scattering rate above the limit set by the ground state hyperfine structure (see below). 

To simulate the optical cycling rate of the Q(1) line of AlF, we solve the optical Bloch equations for the 72-level system, following the notation presented in \cite{Ungar1989, Devlin2018}. The solid red lines in figure \ref{fig:obe} a) represent the excited-state population, $\rho_{ee}$ for the Q(1) line as a function of the laser intensity, normalized to the two-level saturation intensity $I_{\textrm{sat}}=\pi h c \Gamma/(3\lambda^3)=0.93$ W/cm$^2$. The laser frequency is set to the center of the Q(1) line. This is an idealized case, assuming no losses to $v''>0$ to demonstrate that a high scattering rate can be achieved despite the large number of levels involved and despite the unresolved hyperfine structure in the ground state. As expected, with a linearly polarised laser and no polarisation modulation ($\delta_{\textrm{PM}}=0$), the scattering rate peaks at about $\Gamma/30$, limited by the precession of dark superposition states in the ground state. At high intensity, the laser interaction dominates over the hyperfine interaction, and this begins to stabilise the dark states. The excited state fraction falls off, and the system increasingly resembles the fine structure picture described above. The simulations also show that if the polarisation axis rotates about the k-vector of the light at a rate $\delta_\textrm{PM}=\Gamma/2$, the dark states are effectively destabilized, and a high scattering rate is restored. The time-evolution of the population in the ground and excited states is plotted for $I = 8 I_{sat}$ and $\delta_\textrm{PM}=0$ in figure \ref{fig:obe} b), showing that the steady state is reached at $\Gamma t=1000$.

The results from the optical Bloch equations can be compared to the solutions of a rate equation model, which neglects all coherences (dashed, red line in figure \ref{fig:obe} a)). The higher Q lines involve many more levels for which solving the Bloch equations becomes impractical. However, we solve the rate equations for the Q(2) and Q(3) lines shown as dashed green and dashed blue curves, respectively. The figure demonstrates that, for low laser intensities ($s\textcolor{black}{\ll} 1$), the rate-equation model describes the complex multi-level system well. However, for high intensities there are significant differences. The rate equations predict that the scattering rate increases for higher Q lines because the span of the hyperfine structure reduces with increasing $J'$. The number of coherent dark states decreases from $1/3$ to $1/9$ of the total number of hyperfine transitions for the Q(1) to Q(4) lines. It is therefore expected that the peak scattering rate also increases for higher Q lines. 

It is interesting to contrast AlF to the case of TlF, where recently it was shown that unresolved hyperfine structure in the ground state significantly limits the optical cycling rate \cite{grasdijk2021centrex}. Here, the hyperfine structure in the excited state is about 30 times larger than in AlF and the natural linewidth of the optical cycling transition is about 50 times narrower. The number of excited states which are accessible with a single laser is less than a third of the number of ground states, and polarisation modulation alone is insufficient to recover the molecules pumped into dark states. Instead, laser polarisation modulation must be combined with resonant microwave radiation to recover the molecules from the dark states.

\section{Experimental Setup}
\begin{figure}[tb]
\centering
\includegraphics[]{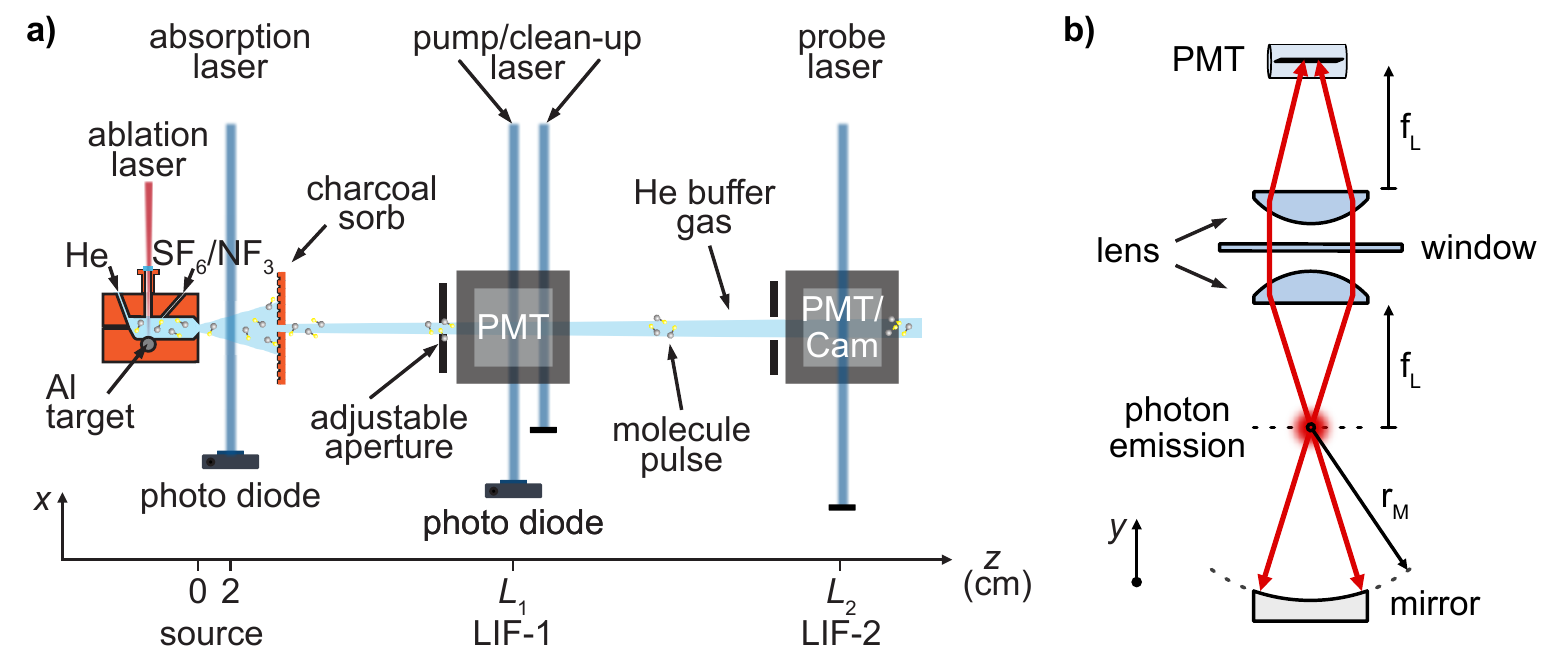}
\caption{\label{fig:BGS schematic}A schematic of the experimental setup. a) The AlF molecules are produced in a cryogenic helium buffer gas cell by laser ablation of aluminium in the presence of SF$_6$ or NF$_3$. The cell is surrounded by two layers of radiation shields held at 30~K and 2.\textcolor{black}{7}~K, respectively. The 2.7~K shield is coated with coconut-derived charcoal which acts as a sorption pump for the helium buffer gas. The molecules pass through an aperture in the shields and can be probed at various locations along the beam. For the cycling and deflection experiments the width of the transverse velocity distribution is reduced by increasing the distance of the detectors to the source and by using an adjustable aperture. b) A 1:1 imaging system is used to collect the laser-induced fluorescence emitted by the molecules and is detected by a UV sensitive PMT or a sCMOS camera outside the vacuum chamber.}
\end{figure}
We use a cryogenic molecular beam source to produce a pulsed molecular beam of AlF molecules and laser absorption, laser induced fluorescence, and two-photon ionisation to detect the molecules. A sketch of the experimental apparatus is shown in figure \ref{fig:BGS schematic}. The source is similar to our previous design \cite{Truppe2018} which is based on the pioneering work presented in \cite{Maxwell2005, Barry2011}. The operation principle of such a cryogenic buffer gas molecular beam has been extensively reviewed in \cite{Hutzler2012}. The output of a pulsed Nd:YAG laser, operated at 1064 nm with a pulse energy of 15-30 mJ, is focused to a spot size of about 0.4 mm onto a rotatable Al target inside a copper cell with external dimensions of (w, h, l) = (35, 35, 45) mm and an internal bore with diameter of 10 mm. To reduce the forward velocity, a double-stage cell configuration can be used \cite{Patterson2007, Lu2011}. The cell is rigidly attached to a closed-cycle cryocooler and cooled to 2.7 K. The vaporized Al atoms collide with a reactant gas, which flows continuously into the cell via a copper tube, insulated from the cell walls. We use either NF$_3$ or SF$_6$ gas at flow rates of 0.001 and 0.03 sccm, respectively. Cryogenic helium buffer gas flows continuously through the cell, cools the molecules and extracts them through a 5 mm aperture. The cell is surrounded by two layers of radiation shields, an inner layer cooled to 2.\textcolor{black}{7} K and an outer layer cooled to 30~K. The 2.\textcolor{black}{7} K shield is covered with coconut-derived charcoal, which acts as a sorption pump for helium and keeps the pressure in the source chamber below 10$^{-7}$ mbar during operation. The apparatus cools to below 3 K in about 4 h and can be heated to room temperature in about 3 h using resistive cartridge heaters with a total power of 150 W. The molecular beam is collimated by a 10 mm and a 15 mm aperture in the inner and outer radiation shields, respectively. About 2 cm from the cell aperture a laser beam intersects the molecular beam at right angles to measure the extraction time, molecule number, their density and transverse velocity distribution. An additional, adjustable aperture, located 4 cm before the center of the first detector (LIF-1), with diameters ranging between 1 and 15 mm can be used to restrict the transverse velocities that enter the optical pump and probe regions. 

We have two identical molecular beam source chambers, with slightly different distances to the fluorescence detectors. Beam machine one is used for the SF$_6$ source characterisation presented in section \ref{sec:buffer} and for the optical cycling experiments presented below. For the source characterisation the distances to the first and second detectors are $L_1=35$ cm and $L_2=63$ cm, respectively. For the optical cycling experiments we reduce the transverse velocity distribution by increasing the distance between the source and the two detectors ($L_1=54$ cm and $L_2=82$ cm). Beam machine two is used to characterise the NF$_3$ source with $L_1=20$ cm and $L_2=44$ cm. 

To generate the UV light needed for our experiments, we use two successive, resonant doubling stages to frequency-quadruple a cw, Ti:sapphire laser. This typically produces 150-250 mW of UV radiation near 227.5 nm ($0-0$ band) and 231.7 nm ($0-1$ band). The fundamental laser frequencies (near IR) are determined with an absolute accuracy of 10 MHz with a calibrated wave meter. The same wave meter is used to lock the laser frequencies to the transition frequencies. The laser beams intersect the molecular beam at right angles and the resulting laser induced fluorescence is imaged onto either a PMT (quantum efficiency of 0.35) or a UV-sensitive sCMOS camera (quantum efficiency of 0.48). The probability to detect an emitted photon is mainly limited by the collection efficiency of the optics, which cover a solid angle of 1 steradian. Including the transmission losses through the vacuum window, lenses and the reflectivity of the mirror gives a total detection efficiency of 0.025 and 0.016 for the two detectors, respectively. A fine stainless steel mesh with a transparency of 80\% can be inserted above and below the molecular beam to apply a homogeneous electric field within the interaction region. This allows studying optical cycling in the presence of dc electric fields. To measure the two-photon ionisation rate of AlF, the LIF detector can be replaced with a Wiley-McLaren type time-of-flight mass-spectrometer in combination with a microchannel plate detector.

\section{{\label{sec:buffer}}Buffer gas molecular beam of AlF}
\begin{figure}
\centering 
\includegraphics[]{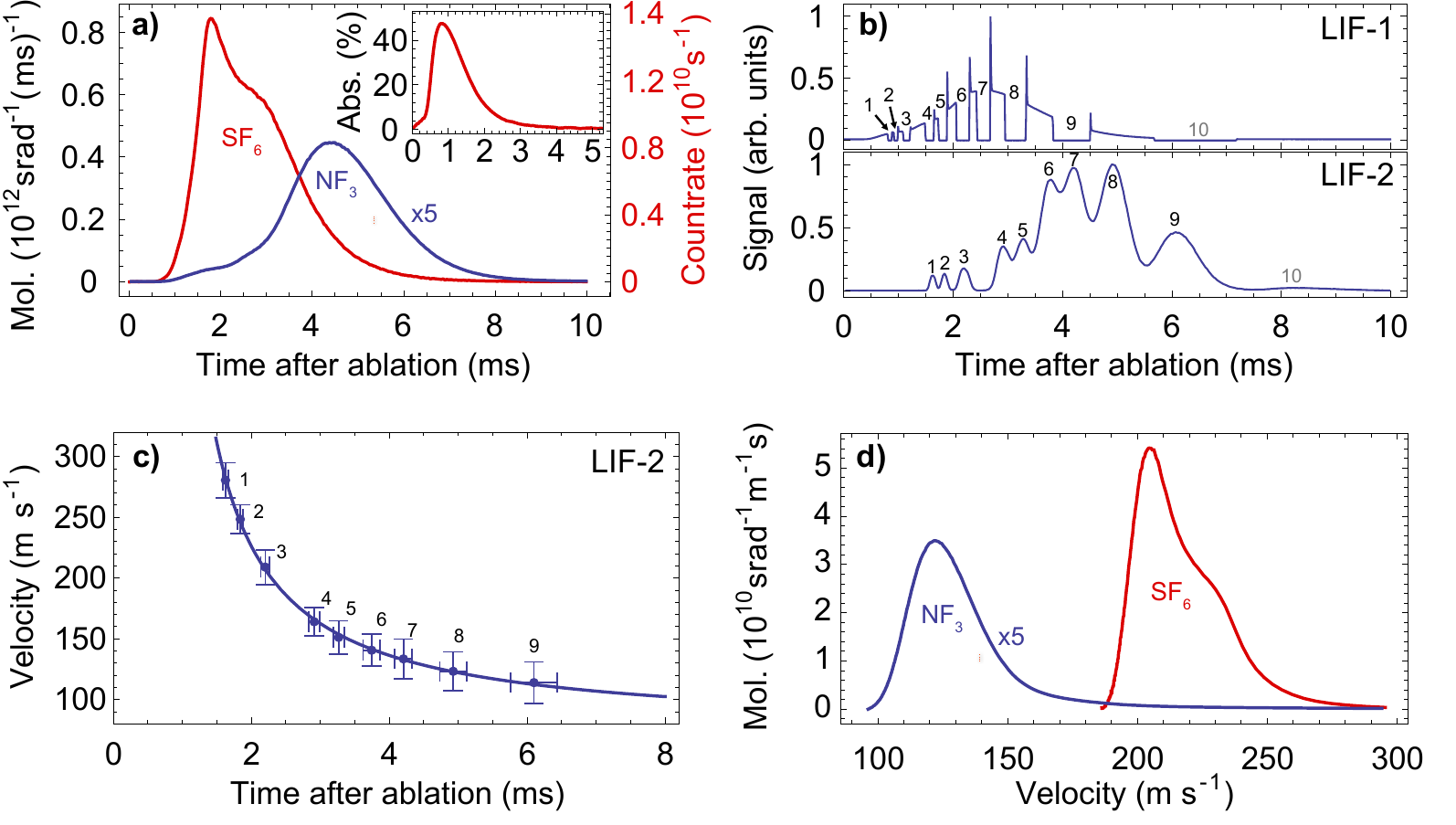}
  \caption{Characterisation of the cryogenic buffer gas molecular beam of AlF using a single (red, \textcolor{black}{SF$_6$}) and double-stage (blue, \textcolor{black}{NF$_3$}) cell. a) time of flight (TOF) fluorescence signal. Inset: laser absorption measured 2 cm downstream from the cell aperture as a function of time. b) - d) Method to determine the velocity distribution with a single laser frequency using a pump-probe scheme on the R(0) line. b) A rapidly shuttered high-power laser beam chops the TOF profile in LIF-1 into narrow temporal segments by pumping molecules to the dark $J''=2$ level (top). A low power laser beam in LIF-2 probes the population in $J''=0$ (bottom). Each trough in the top panel (molecules are not pumped) has a corresponding peak in the bottom panel. The time-difference between the center of the trough and the center of the peak determines the mean velocity of the molecules within this segment.  c) Mean velocity of each segment (data points) and FWHM of the velocity distribution (vertical error bars) as a function of arrival time with fit (solid curve, $v(t)=a+b/(t-t_0)$). d) TOF profile converted to velocity distribution using the fit function from c).}
\label{fig:velocity-measurement}
\end{figure}
The properties of the molecular beam are characterised using the non-cycling R(0) rotational line of the $A^1\Pi,v'=0 \leftarrow X^1\Sigma^+,v''=0$ band. We probe the density of the molecules in the SF$_6$ beam via absorption spectroscopy 2 cm from the cell aperture as well as 35 cm from the cell via laser induced fluorescence and absorption. Figure \ref{fig:velocity-measurement} a) shows a typical time-of-flight profile of AlF for two distinctively different sources measured in the two different beam machines. The one shown in red results from a single-stage cell using SF$_6$ as fluorine donor gas with a distance of $L_1=35$ cm from the source\textcolor{black}{, optimized to maximize the beam brightness}. In blue we show a molecular beam emerging from a double-stage cell using NF$_3$\textcolor{black}{, optimized for a low and consistent forward velocity}. The distance to LIF-1 is set to $L_1=42$ cm from the cell aperture. The detection laser is locked to the center of the spectral line and the laser power is chosen to saturate the fluorescence. The inset depicts the temporal absorption profile for the SF$_6$ beam, recorded 2 cm from the cell aperture, with the frequency of the detection laser locked to the peak of the Doppler broadened resonance (1 GHz FWHM) and a power of 0.1~mW ($I\approx 10^{-3} I_{\textrm{sat}}$). \textcolor{black}{The low laser intensity prevents pumping the molecules to dark rotational states during their interaction with the laser beam and ensures uniform absorption.} The transverse velocity spread of the molecules leaving the cell is measured to be 200 m/s, which corresponds to an angular spread of the molecular beam of 50$^\circ$ FWHM ($\approx 0.7$ steradian). The number of molecules leaving the cell in the rotational ground state can be determined by the time-integrated resonant-absorption to $N=\frac{A_b v_z}{L \sigma_D}\int \ln{(P_i/P_t)}\text{d}t$, where $A_b$ is the cross-sectional area of the molecular beam at the position of the absorption laser, $v_z$ is the forward velocity and $L$ is the absorption length. \textcolor{black}{To find the peak absorption cross-section, we model the absorption spectrum using the spectroscopic constants from \cite{Truppe2019} together with Voigt profiles, whose Lorentzian contribution is $84$ MHz and whose Gaussian contribution is used to fit to the experimental spectrum. The result is a Doppler-broadened peak absorption cross-section of $\sigma_D=1.6\times 10^{-11}$ cm$^2$ for the R(0) rotational line.} $P_i$ and $P_t$ are the incident and transmitted laser power, respectively. For the SF$_6$ beam this amounts to $N=2.\textcolor{black}{5}\times 10^{12}$ molecules per steradian per molecular pulse in the rotational ground state, which compares well to \textit{atomic} cryogenic buffer gas beams \textcolor{black}{of Yb}, which typically contain about $10^{12}-10^{13}$ atoms per pulse \cite{Patterson2007, Hutzler2012}. \textcolor{black}{Assuming that a comparable number of Al atoms is produced in our cell, it is clear that the reaction of hot Al with NF$_3$ or SF$_6$ to form AlF is very efficient.}  

Absorption measurements close to the cell aperture to determine the density of the molecules can be misleading due to collisions with the high density of helium buffer gas in this region. To measure the molecular beam brightness further downstream we carefully characterise  the fluorescence detector. Here, the transverse velocities of the molecular beam can be restricted using a variable aperture and a camera is used to measure the spatial distribution of the molecular fluorescence for the different apertures. This allows us to calibrate the imaging system and compare it to the results of ray-tracing software\textcolor{black}{\footnote{\textcolor{black}{OSLO EDU, Lambda Research}}}. The total collection efficiency of the optics, including the transmission through the lenses and window amounts to $\epsilon=0.07(1)$. The quantum efficiency of the PMT is $\epsilon_{\text{qe}}=0.35(5)$. The detected solid angle of the molecular beam is determined by a 15 mm diameter aperture that restricts the molecular beam and the measured laser beam diameter ($e^{-2}$) of 2 mm to $\Omega=2.4(5)\times10^{-4}$. The beam brightness is $\Phi=N_\text{c}/(\epsilon\epsilon_{\text{qr}}\Omega \langle n^\infty_{\text{ph},R}\rangle)=1.7(4)\times 10^{12}$ molecules per steradian per pulse in the rotational ground state, where $N_c$ is the total number of photon counts per pulse, and $\langle n^\infty_{\text{ph},R}(J''=0)\rangle=3$. \textcolor{black}{When the fluorescence saturates, the average number of photons emitted by the molecules is independent of their forward velocity (see sections \ref{sec:rotBranching} and \ref{sec:lif})}. The molecular beam brightness, measured by fluorescence agrees well with the absorption measurement close to the source. Additionally, we verify this measurement by restricting the molecular beam using an aperture with a diameter smaller than the spot-size of the excitation laser. This defines the solid angle more precisely and makes sure that all molecules experience the same laser intensity. We also measure the absorption of the probe laser in LIF-1 to $P_i/P_{t,min}\simeq 4(1)\times10^{-3}$ (peak absorption). The calculated peak absorption cross-section including the hyperfine structure and the residual Doppler broadening of 40 MHz is then $\sigma=5 \times 10^{-11}$ cm$^2$. The peak absorption corresponds to a peak density of $5\times10^7$ cm$^{-3}$ and about $5\times10^8$ molecules within the anticipated capture volume of the MOT. The time integral of the absorption profile allows converting the absorption signal into a beam brightness of $1.5\times 10^{12}$ molecules per steradian per pulse in the rotational ground state, consistent with the fluorescence measurement.

The large natural linewidth and the hyperfine structure in the excited state preclude using Doppler-sensitive LIF detection to determine the forward velocity distribution of the molecular beam. Instead, we use an optical pump-probe scheme in combination with time-of-flight measurements. A high power laser beam ($\approx 5$ W/cm$^2$), tuned to the R(0) line, intersects the molecular beam in LIF-1 and pumps $>90\%$ of the interacting molecules to $J''=2$. Further downstream, in LIF-2, the same laser probes the population left behind in $J''=0$ with a ten times lower intensity. By rapidly shuttering the pump laser using an acousto-optical modulator we divide a single time-of-flight profile into ten temporal segments and measure the velocity of the molecules within each segment via their time-of-flight to the second LIF detector 22 cm further downstream (figure \ref{fig:velocity-measurement} b). The spikes in the top panel of figure \ref{fig:velocity-measurement} b) are caused by molecules that are already present in the detection volume when the laser light is rapidly turned on. This simple technique allows us to record the forward velocity as a function of the arrival time in a single shot (figure \ref{fig:velocity-measurement} c) and provides a fast and accurate alternative to Doppler-sensitive LIF detection. A fit to the data is used to convert the time-of-flight profile from panel a) into a velocity distribution, shown in panel d). 

If both sources are operated under identical conditions, the forward velocity for the NF$_3$ beam is typically 10-15\% lower and the brightness is the same within a factor 2. However, when using SF$_6$ as a fluorine donor, we observe a steady increase of the molecular beam velocity with the number of ablation pulses. This is most likely caused by a build-up of fluoride and sulphur compounds (ablation dust) inside the buffer gas cell. The dust prevents the helium buffer gas from thermalizing with the cell walls after being heated by the ablation plasma, which results in a faster and hotter molecular beam. Cleaning the cell restores a slow beam. Flowing He and SF$_6$ for a day through a clean cell without the ablation laser running results in a slow beam, indicating that this effect is not caused by SF$_6$ ice or a poor helium pumping. These issues can be mitigated by using a cell with a larger volume, which is less sensitive to dust build-up and heating effects, but produces longer, less dense molecular pulses. 

When NF$_3$ is used instead of SF$_6$, the molecular beam has a lower forward velocity (170 m/s, single stage cell and 130 m/s double-stage) and does not speed up over time. The number of AlF molecules produced this way is equivalent and saturates at a much lower flow rate of the fluorine donor gas. The SF$_6$ beam delivers more molecules to the detector due to the lower divergence and slightly higher forward velocity. This beam is adequate for spectroscopy and optical cycling experiments. However, for loading a MOT a more stable beam with a lower forward velocity might be more advantageous.

\section{Optical cycling}
The optical cycling rate is measured in three different ways. First, we compare the LIF signal from the cycling Q lines with the LIF signal when driving the non-cycling P or R lines. Second, we determine the rate at which the ground-state molecules are optically pumped into the dark $X^1\Sigma^+,v''=1$ state. Third, we measure the radiation pressure force via molecular beam deflection.
\subsection{\label{sec:lif}Calibration of the laser-induced fluorescence}
\begin{figure}
\centering
\includegraphics[]{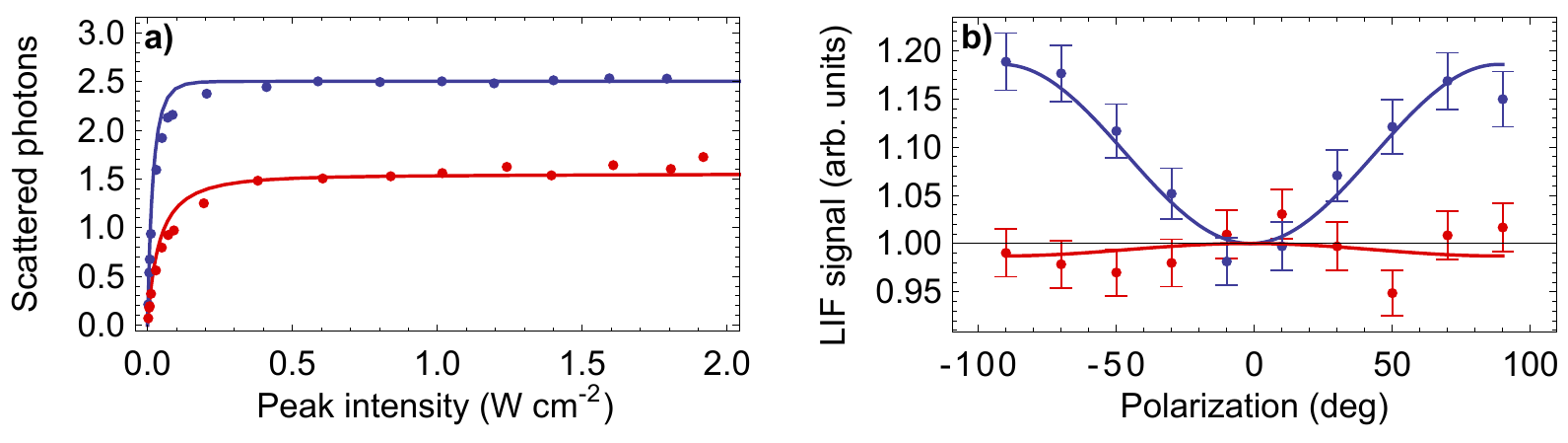}
\caption{\label{fig:r1-p3-plots}
Fluorescence yield of non-cycling transitions. a) The fluorescence of the R(1) (blue data points) and the P(3) (red data points) lines saturates at 2.5 and 1.5 photons per molecule, respectively. Simulations based on a rate equation model are shown as solid curves. b) Rotating the polarisation of the excitation laser changes the detected fluorescence intensity for the R(1) line (blue) significantly, but not for the P(3) line (red). Only if the polarisation is set along $y$, i.e. along the axis of the detector \textcolor{black}{($\Theta=0$)}, the ratio of the signals from the P and R lines is equal to the ratio predicted by the Hönl-London factors (shown in panel a)).}

\end{figure}
\begin{figure}
\centering
\includegraphics[]{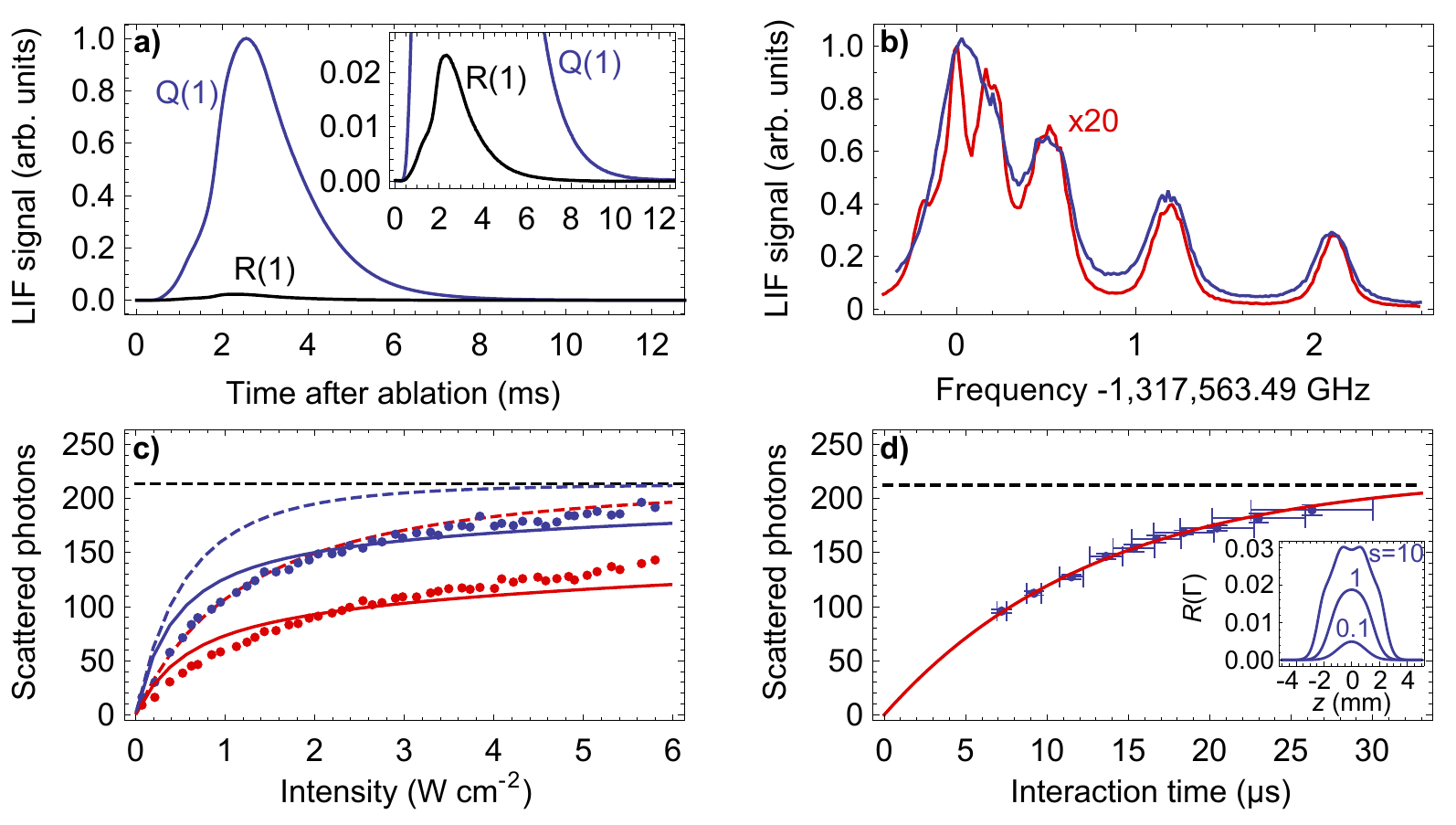}
\caption{\label{fig:q-saturation-plots}
Optical cycling on Q lines determined by calibrating the laser induced fluorescence with a standard candle. a) Compared to a non-cycling transition (R(1), red), the LIF signal is strongly enhanced by using a cycling transition (Q(1), blue) to probe the molecular beam. b) A Q-branch spectrum recorded using a low (blue) and high (red) laser intensity demonstrating that the molecules cycle on all Q lines. c) Saturation of the Q(1) fluorescence for molecules travelling with $v_z=150$ (blue points) and $v_z=300$ m/s (red points). Simulations for each velocity based on optical Bloch equations and rate equations are shown as solid curves and dashed curves, respectively. d) Number of photons scattered on the Q(1) line as a function of the interaction time $t_i=w_z/v_z$ with the laser, where $w_z=3.7$ mm is the $e^{-2}$ diameter of the laser beam along $z$. The solid curve is a fit to the data using equation \ref{eq:nphot}. Inset: the simulated photon scattering rate as a function of $z$ for $w_z=3.7$ mm and three values of the saturation parameter $s$ in the absence of losses to $v''=1$.}
\end{figure}

A straightforward way to determine the number of optical cycles is to compare the fluorescence yield from a cycling transition to a standard candle. For the cycling Q lines, the average number of photons that can be scattered is determined by the vibrational branching ratio to $\langle n^\infty_{\textrm{ph},Q}\rangle=213(30)$ (see section \ref{sec:vibBranching}). All other potential loss channels can be neglected at this level and are summarized in our previous study \cite{Truppe2019}. To restrict off-axis excitation, the molecular beam is collimated with a round, 2 mm diameter aperture placed 4 cm before the first LIF detector. For the data presented here, the laser beam is collimated to a $e^{-2}$ diameter of $w_z\textcolor{black}{=}\SI{3.7}{mm}$ and $w_y=\SI{2.9}{mm}$. This ensures sufficient intensity is available to saturate the transition, that the transverse extent of the laser exceeds that of the molecular beam, and that the interaction volume can be reliably imaged onto the PMT. 

First, we perform a consistency check to make sure that the standard candles are indeed a good reference to calibrate the fluorescence of the cycling transitions. Figure \ref{fig:r1-p3-plots} a) shows the laser induced fluorescence for the R(1) and P(3) line, time-integrated over the duration of the molecular pulse, as a function of the laser intensity. The ratio of the two limiting values is fixed by the ratio of the Hönl-London factors. A deviation from this ratio would lead to a systematic error in the calibration. The solid curves are the result of a theoretical simulation of the light-molecule interaction using rate equations, the measured velocity distribution and spatial profile of the molecular and laser beam. The P line saturates at 1.5 photons and not at $\langle n^\infty_{\textrm{ph},P}\rangle=5/3$ as predicted in section \ref{sec:rotBranching}. This is due to optical pumping of molecules into the dark $M_{F''}=\pm 6$ Zeeman sublevels lowering the average number of photons scattered per molecule slightly. A Q-branch spectrum allows us to normalize the LIF signals to the respective rotational state population in $J''=1$ and $J''=3$. We display the saturation of the R(1) and P(3) line because they both reach the same excited state. This is important, because the excited state hyperfine structure can influence the detected fluorescence yield in subtle ways (see below). In saturation, the two LIF signals indeed reflect the predicted ratio by the Hönl-London factors.

The excited state hyperfine structure can lead to subtle interference effects between decays from the different excited states. The relative weight of $\sigma^+$, $\sigma^-$ and $\pi$ polarised radiation emitted by the molecules depends on the angle of the polarisation of the excitation light with respect to the detector. The fluorescence intensity of $\pi$ polarised light is $I_\pi\propto \sin^2\Theta$ and for $\sigma$ polarised light is $I_\sigma\propto (1+\cos^2\Theta)/2$, where $\Theta$ is the angle between the $\mathbf{k}$-vector of the fluorescence light and the polarisation vector of the excitation light. The position of the fluorescence detector is fixed at a right angle to the $\mathbf{k}$ vector of the excitation light. By averaging over all possible decay channels we expect a periodic modulation of the detected fluorescence with an amplitude of 0.18 for the R(1) and -0.03 for the P(3) line. Figure \ref{fig:r1-p3-plots} b) summarizes this effect. The points with errorbars are the measured fluorescence normalized to their value for $\Theta=0$ and the solid curves are the predicted modulation by calculating the weighted sum over all possible decay channels. Therefore, it is important to choose $\Theta=0$, where the ratio of the two fluorescence signals in figure \ref{fig:r1-p3-plots} a) corresponds to the ratio of the Hönl-London factors. The same is true for the Q(1) line, whose periodic modulation has an amplitude of 0.11, slightly less compared to the R(1) line. 

Figure \ref{fig:q-saturation-plots} a) shows two time of flight profiles of the molecular pulse with the laser tuned to the Q(1) (blue) and R(1) (black) lines. Both traces are taken with the same laser intensity to demonstrate the signal enhancement due to optical cycling. The time-integrated fluorescence signal as a function of the laser frequency for low and high laser intensity is shown in figure \ref{fig:q-saturation-plots} b). The low intensity spectrum is scaled by a factor of 20. Each Q-line shows a similar signal enhancement at high laser power, demonstrating that they can all cycle effectively.

The saturation of the Q(1) fluorescence for two forward velocities $v_z=150$ m/s (blue) and $v_z=300$ m/s (red), is shown in figure \ref{fig:q-saturation-plots} c). The solid curves are simulations based on the optical Bloch equations and the dashed lines are the predictions from the rate equation model. We use the measured laser beam profile and forward velocity to predict $\textcolor{black}{N}=\int R(t) dt$ for each theoretical model and equation \ref{eq:nphot} to \textcolor{black}{predict $\langle n_{\textrm{ph}}\rangle$}. It is clear that a high optical cycling rate can be obtained despite the large number of hyperfine components involved in the cycling scheme. The rate equations predict that the fluorescence saturates at a lower intensity because they do not account for the effects of coherent dark states. Figure \ref{fig:q-saturation-plots} d) shows the number of photons scattered as a function of the interaction time, defined by $t_i=w_z/v_z$, where $w_z$ is the $e^{-2}$ diameter of the laser beam. The solid curve is a fit to the data using equation \ref{eq:nphot} with $R$ as fit parameter and $r=0.9953$ the theoretical branching ratio. The fit gives $R=17.2(2)\times 10^6$ s$^{-1}$ for the effective scattering rate for a saturation parameter $s=6$. We estimate the systematic uncertainty in determining the number of photons scattered to be 10\%. This is mainly due to uncertainties in the calibration procedure that arise from a non-linearity in the PMT gain, combined with variations in the source flux and velocity distribution. 

\subsection{{\label{sec:optical pumping}}Optical pumping into $X^1\Sigma^+,v=1$}

\begin{figure}[tb]
\centering
\includegraphics[]{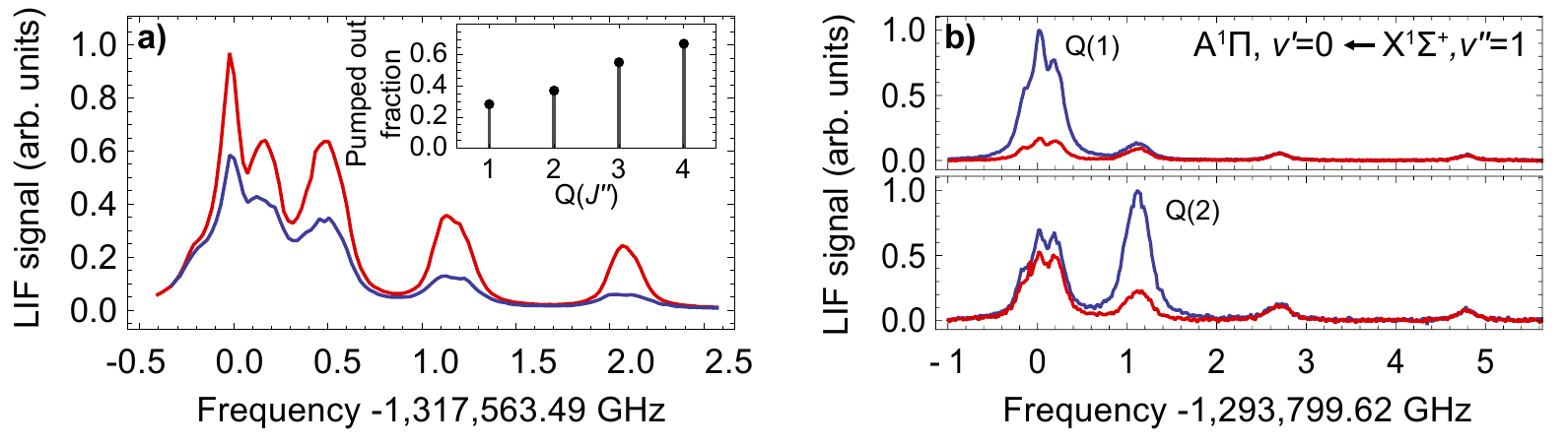}
\caption{\label{fig:qlines} a) The depletion of the $v''=0$ population by a strong pump laser in LIF-1 is probed in LIF-2 with a low-intensity laser beam that has the same frequency as the pump laser. The spectrum in blue (red) is with the pump laser in LIF-1 on (off). Inset: fraction of molecules pumped from $v''=0$ to $v''=1$. b) Optical pumping to $X^1\Sigma^+, v''=1$ with the pump laser tuned to the Q(1) (top) and Q(2) (bottom) line of the $0-0$ band. The $v''=1$ population is probed by recording a Q-branch spectrum of the $A^1\Pi, v'=0 \leftarrow X^1\Sigma^+, v''=1$ band in LIF-2 with (blue) and without (red) the pump laser present in LIF-1.}
\end{figure}

The second method to determine the photon-scattering rate is to measure the rate at which the molecules are pumped out of $X^1\Sigma^+,v''=0$ and into the dark $X^1\Sigma^+,v''=1$ level. For this we use a high-intensity laser beam ($\approx 20$ W/cm$^2$, $w_z=1.3$ mm, $w_y=1.1$~mm) tuned to the $A^1\Pi, v'=0\leftarrow X^1\Sigma^+,v''=0$ transition. To probe the molecules remaining in $X^1\Sigma^+,v''=0$ we split off a small fraction of the pump laser beam and intersect it with the molecular beam a second time in LIF-2. Figure \ref{fig:qlines} a) shows a Q-branch spectrum recorded in LIF-2 with the pump laser present (blocked) in LIF-1 in blue (red). The fraction of the $v''=0$ population that is pumped out by the high-intensity laser is 29\%, 37\%, 55\% and 67\% for the Q(1) to Q(4) lines, respectively. The data indicates that higher Q lines scatter faster at the same laser intensity and therefore pump out more molecules for the same interaction time. This is consistent with the predictions of the rate equation model presented in figure \ref{fig:obe}. 

The number of photons scattered by the molecules is related to the population remaining in $v''=0$ by $\langle n_{\textrm{ph}}\rangle=-\ln (P_{v''=0})/(1-r)=73(11)$, 98(15), 170(25) and 236(35), for the Q(1) through Q(4) lines, respectively (see section \ref{sec:vibBranching}). The pumped-out fraction for higher rotational lines is slightly overestimated because the molecules experience a radiation pressure force which leads to a Doppler shift in LIF-2 of about 17 MHz per 100~photons scattered (see next section). Additionally, the molecules are displaced from the center of the detection region which decreases the detection efficiency slightly. The number of photons scattered on the Q(1) line corresponds to an effective scattering rate of $R=16(2)\times10^6$ s$^{-1}$ which increases to $R=42(7)\times10^6$ s$^{-1}$ for the Q(4) line. This is in good agreement with the scattering rate determined in the previous section. We have not performed optical Bloch equation simulations for the high Q-lines, but we expect that the peak scattering rate also increases for higher Q lines. In addition, the higher Q lines reach the peak scattering rate at a lower intensity or shorter interaction time which leads to an effectively higher scattering rate for the same laser-beam diameter.

To verify that the molecules are indeed pumped to $X^1\Sigma^+,v''=1$, we probe the $v''=1$ population in LIF-2. Figure \ref{fig:qlines} b) shows Q-branch spectra of the $A^1\Pi,v'=0\leftarrow X^1\Sigma^+,v''=1$ band using a second UV laser tuned to 231.7~nm with an output power of up to 150~mW. The top panel shows the $0-1$ spectrum with (without) the $0-0$ pump laser present in LIF-1 in blue (red). The  top panel shows the $0-1$ spectrum with the pump laser tuned to the Q(1) line. The bottom panel is the same spectrum, but with the pump laser tuned to the Q(2) line. The total number of $v''=1$ molecules that are produced in the source relative to the number of molecules in $v''=0$ is typically $9(2)\%$. The increase in $v''=1$ population (blue curves) corresponds to a fraction of molecules that is pumped out of $v''=0$ of $36(8)\%$ and $31(7)\%$ for the Q(1) and Q(2) lines, respectively. This is consistent with the pumped-out fraction shown in the inset of Figure \ref{fig:qlines} a). The uncertainty is dominated by fluctuations in the $v''=1$ population relative to $v''=0$.

\subsection{Deflection of the molecular beam by radiation pressure}
\begin{figure}[tb]
\centering
\includegraphics[]{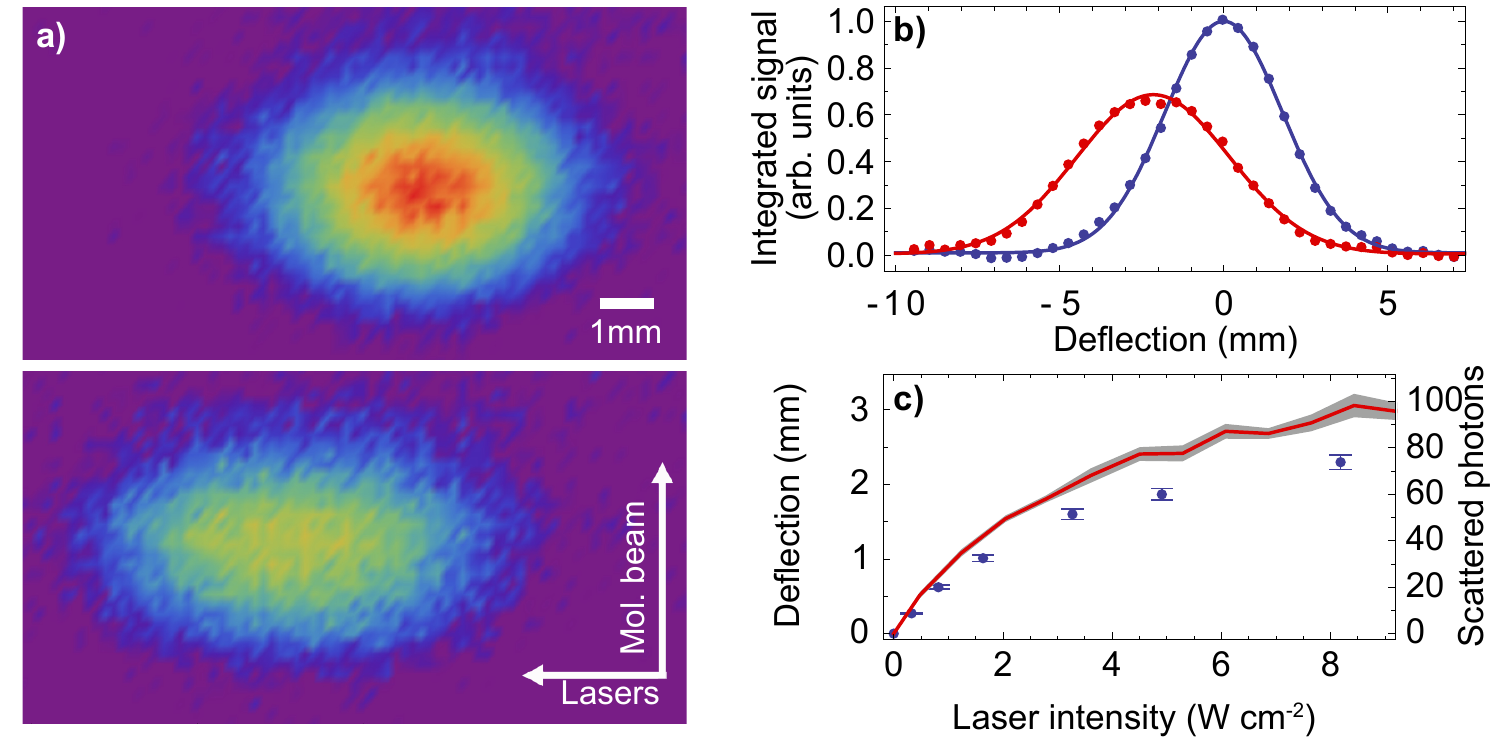}
\caption{\label{fig:deflection}Deflection of the molecular beam by radiation pressure from a high-intensity pump laser tuned to the Q(2) rotational line of AlF (in LIF-1). A clean-up beam recovers population from $v''=1$. A fraction of the pump laser beam is split off and used to probe the molecules in LIF-2.  a) Image of the molecular beam passing through the probe laser beam with the pump laser beam off (top) and on (bottom). b) A Gaussian fit to the integrated image of the molecular beam yields a deflection of $d=2.15$ mm. c) Deflection and corresponding number of photons scattered as a function of the laser intensity. The red curve is a trajectory simulation using the rate-equation model to calculate the radiation pressure force exerted on the molecules. As expected, the rate-equation model overestimates the number of scattered photons slightly \textcolor{black}{(see section \ref{sec:theory})}.}
\end{figure}
After establishing an optical cycling scheme, the next step towards laser slowing and magneto-optical trapping is to demonstrate a measurable radiation pressure on the molecular beam. This method is also a very direct way to determine the number of photons scattered by the molecules. Robert Frisch, motivated by Otto Stern's proposal \cite{Stern1926}, has demonstrated the deflection of an atomic beam using resonance radiation already in 1933 \cite{Frisch1933}. This was followed by many more experiments using lasers in combination with a variety of atomic and molecular beams \cite{Picque1972,Bjorkholm1980, Shuman2009, Chen2017, McNally2020}.

Here, we use a high-intensity (10 W/cm$^2$, $w_z=0.9$ mm and $w_y=2$ mm) laser tuned to the Q(2) rotational line to exert radiation pressure onto the molecular beam and measure the transverse deflection by imaging the molecular fluorescence onto an sCMOS camera. A round 2 mm aperture located in front of the pump region in LIF-1 collimates the molecular beam transversely to 1 m/s (FWHM). This guarantees that the entire molecular beam can be imaged onto the camera. The imaging system is calibrated by translating the collimating aperture in LIF-1 along $x$ and measuring the average displacement of the molecular fluorescence on the camera. A clean-up beam (100 mW in $w_z=w_y=2$ mm) intersects the molecular beam a few cm downstream from the pumping region (see figure \ref{fig:BGS schematic}) and recovers $>95\%$ of the molecules lost to $v''=1$. The $v''=0$ molecules are detected in LIF-2 by splitting off a fraction of the $0-0$ pump light and directing it through LIF-2.

The short excited state lifetime and small Franck-Condon factor of the $0-1$ transition requires a high laser intensity to saturate the transition. For future laser-slowing and cooling experiments it is important to reach a repumping rate that exceeds the rate at which molecules are pumped into $v''=1$. We measure a repumping rate of $\approx 10^6$ s$^{-1}$, fast enough to guarantee efficient repumping. The clean-up beam pumps also the small population of $v''=1$ molecules produced in the source to $v''=0$ which is then detected in LIF-2 by the probe beam. We measure this fractional increase in $v''=0$ population by turning off the pump laser and determining the difference in LIF signal with and without the clean-up beam. The difference is subtracted from the image of the deflected beam because it originates from molecules that are not deflected by the radiation pressure force in LIF-1.

By absorbing a photon, the molecule acquires a linear momentum $\mathbf{p}=\hbar \mathbf{k}$ in the direction of this photon. The resulting deflection of the molecular beam, measured in LIF-2, is $d=\langle n_{\textrm{ph}}\rangle v_r L/v_z$, where $v_r=\hbar k/m = 3.8$ cm/s is the recoil velocity per photon, $L=28$ cm is the distance between the pump and probe regions and $v_z=270(20)$ m/s is the forward velocity of the molecules. 

Figure \ref{fig:deflection} summarizes the results of this experiment. Panel a) shows a binned camera image with the pump laser turned off (top) and turned on (bottom). The image is integrated and fitted with a Gaussian to determine the average displacement of $d=2.15$ mm (panel b), indicating that the molecules scatter $\langle n_{\textrm{ph}}\rangle=55(4)$ photons. This is consistent with the value derived from optical pumping, given the shorter interaction time due to the smaller waist diameter $w_z$. The recoil velocity is comparable to the width of the transverse velocity distribution of the molecular beam which leads to a broadening of the molecular beam along $x$. This is a result of the random direction of the spontaneously emitted photons which leads to velocity diffusion and therefore a symmetric broadening of the molecular beam. The diffusion coefficient can be approximated by $D\approx\left(\hbar k/m\right)^2 R$, with $R$ being the photon scattering rate. This momentum diffusion gives rise to a characteristic velocity change of the molecules of $R/k=0.54$ m/s \cite{Letokhov1981}, and therefore broadens the transverse velocity distribution by about 54\%. A more accurate approximation is given in \cite{Bjorkholm1980}. The upper bound for the transverse velocity after scattering $\langle n_{\textrm{ph}}\rangle=55$ photons is $v_{t}=2(v_z\Delta\theta+\sqrt{\langle n_{\textrm{ph}}\rangle/3}h/(m\lambda))=1.33$~m/s, adding 0.33~m/s to the 1~m/s of the original beam. Here $\Delta\theta=1.9$~mrad is the half-angular divergence of the molecular beam, before the interaction with the laser. We measure a 33\% increase in the width of the molecular beam, consistent with the simple model presented above. The symmetric broadening in figure \ref{fig:deflection} b) indicates that the scattering rate along $y$ is homogeneous. This is surprising since the laser beam $w_y=2$~mm has the same diameter as the aperture that is used to collimate the molecular beam. However, the inset of figure \ref{fig:q-saturation-plots} d) shows that for high laser intensities ($s\geq10$), the Gaussian wings contribute significantly to the overall scattering rate.

The area under both curves in figure \ref{fig:deflection} b) is identical, showing that the total number of molecules is conserved and verifying the effect of the repump laser. The Doppler shifts induced by the deflection are negligible because of the large transition linewidth. Figure \ref{fig:deflection} c) shows the deflection and the inferred average number of scattered photons for a slightly faster molecular beam ($v_z=330$~m/s) as a function of the laser intensity. The red solid curve is the result of a Monte Carlo trajectory simulation using the rate equation model to predict that scattering rate, the measured laser beam profile, velocity distribution and spatial distribution of the molecular beam as input parameters. The simulation results fit the data well, which shows that the rate equations predict the scattering rate well for short interaction times.

The mean scattering rate of the molecules, averaged over the interaction volume is $R=23(4)\times 10^6$ s$^{-1}$, which corresponds to an acceleration of $a=v_r R=8.7(1.5)\times 10^5$~m/s$^2$.

To slow the molecular beam to rest using radiation pressure, the leak to $v''=1$ must be closed with a repump laser, which allows scattering about 10$^4$ photons. This typically lowers the scattering rate by at least a factor of two due to the additional ground-state levels that couple to the same excited state. However, the molecules also reach a higher peak scattering rate as they are not pumped to $v''=1$ by the low intensity Gaussian wings of the laser beam. We estimate the stopping distance for the SF$_6$ and NF$_3$ beam presented in figure \ref{fig:velocity-measurement}, conservatively to 6 and 2 cm, respectively. By modulating the polarisation of the $0-0$ light in a slowing configuration the peak scattering rate could be quadrupled, reducing the stopping distance to below 1 cm. Conveniently, the large hyperfine structure of the Q lines covers the Doppler width of the molecular beam.

\section{Parity mixing and two-photon ionisation}

\begin{figure}[tb]
\centering
\includegraphics[]{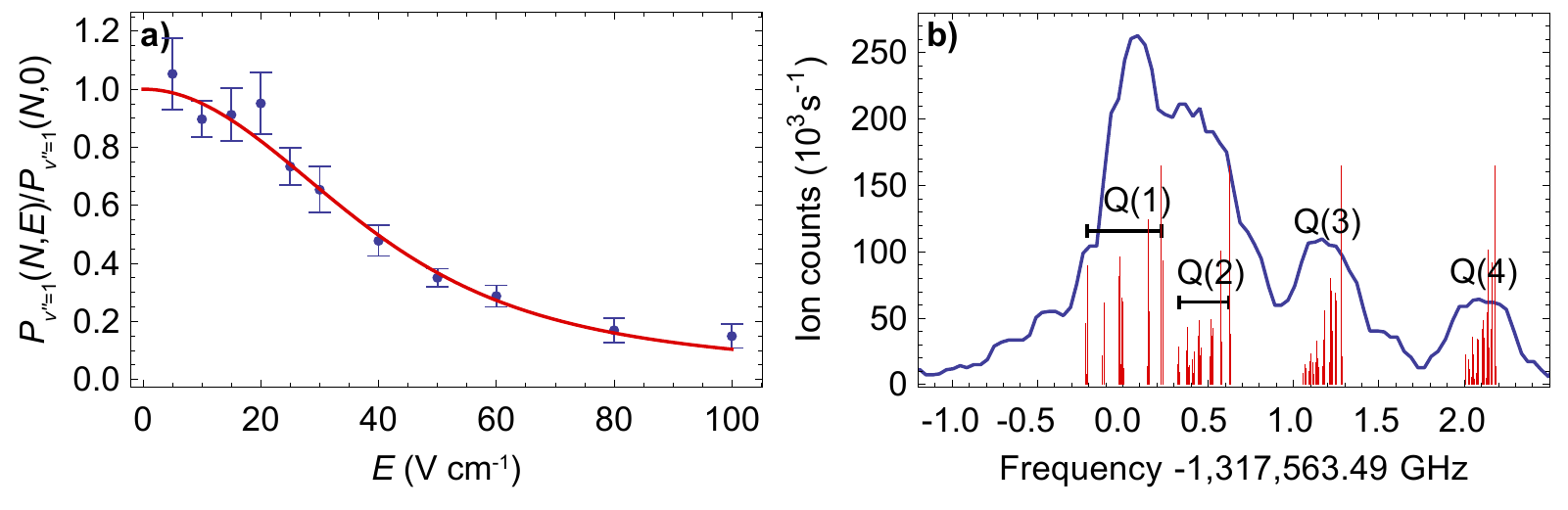}
\caption{\label{fig:losschannels}Rotational loss channels induced by stray electric fields and two-photon ionisation can limit the photon scattering rate. a) Parity mixing in an external electric field opens a loss channel comparable to vibrational branching. \textcolor{black}{The $v''=1$ population, $P_{v''=1}(N,E)$, is measured as a function of the applied electric field $E$ and normalized to the value at zero electric field $P_{v''=1}(N,0)$.} b) Q-branch spectrum up to $J=4$  by detecting the molecular ions in a Wiley–McLaren time‐of‐flight detector. For efficient two-photon ionisation a large excited state population is required. Therefore, a very low ion-extraction field must be used to prevent optical pumping into a rotational dark state via parity mixing.}
\end{figure}

A comprehensive summary of the possible loss channels from the optical cycle is given in our previous papers \cite{Truppe2019, Doppelbauer2020}. In this section we present measurements of two specific loss channels that have not been investigated in detail before, i.e. two-photon ionisation and rotational branching due to parity mixing in the $A^1\Pi, v'=0$ state.

\subsection{Rotational branching due to parity mixing}
A small electric field can mix the closely spaced, opposite parity $\Lambda$-doublet levels in the $A^1\Pi$ state, and therefore break the parity selection rule of an electric dipole transition. After scattering only a few photons, the molecules are pumped into dark rotational states in the ground state and are lost from the optical cycle.

The strongest mixing in the $A^1\Pi, J'=1$ level occurs between the opposite parity $M_{F'}=4$ sublevels of $F'=4$, whose energy-level spacing is only a few MHz (see figure \ref{fig:qbranchhfs}). For very low electric fields, i.e. a small Stark shift compared to the energy-level spacing, the fraction of the opposite parity that is mixed into a given parity level increases quadratically with the electric field, transitions to a linear increase for higher fields and converges to an equal mixture at high electric fields. Given the electric dipole moment of 1.45 Debye, an electric field of a few V/cm is sufficient to open a significant loss-channel to the optical cycle. In the $X^1\Sigma^+$ state the opposite parity levels are separated by at least twice the rotational constant, and parity mixing can be neglected.

When exciting on the Q(1) line, the parity mixing in the the $A^1\Pi, J'=1$ level leads to rotational branching to the dark $J''=0$ and $J''=2$ rotational levels in the electronic ground state. This effect reduces the average number of photons that can be scattered from $\langle n^\infty_{\textrm{ph},Q}\rangle=1/(1-\tau_0 A_{00})$ to a limiting value in high electric fields of only two.

To quantify this potential loss channel, we install electrodes in the optical pumping region (LIF-1), apply a uniform electric field and measure the fraction of molecules that is pumped into $X^1\Sigma^+,v''=1$. The frequency of the pump laser is locked to the center of the Q(1) line and the population in $v''=1$ is probed in LIF-2, similar to the experiment described in section \ref{sec:optical pumping}. Figure \ref{fig:losschannels} a) shows that the fraction of molecules pumped into the $v''=1$ level decreases rapidly with increasing electric field. Following the model described in section \ref{sec:vibBranching} the population in the $v''=1$ level after $N$ optical cycles is given by 
\begin{align}
P_{v''=1}(N,E)&= \tau_0 A_{01} \sum^{N-1}_{n=0}\Big(r-\gamma(1-\tau_0 A_{00}) E^2\Big)^n \simeq \frac{1}{1+\gamma E^2}\Big(1-e^{-N(1-\tau_0 A_{00})(1+\gamma E^2)}\Big),
\end{align}
where $\tau_0 A_{01}\approx (1-\tau_0 A_{00})$. We express the losses from the optical cycle induced by parity mixing as $\gamma (1 - \tau_0 A_{00}) E^2$ and assume a  quadratic dependence with the electric field $E$. Assuming $N=55 \pm 15$ we fit the model using $\gamma$ as a fit parameter. The model fits the data well for $\gamma=\SI{0.0034(6)}{\square\centi\meter\per\square\volt}$. \textcolor{black}{The losses due to vibrational branching and due to parity mixing become equal when $\gamma E^2=1$, which occurs at \SI{17(2)}{\volt\per\centi\meter}. Only if the parity and vibrational losses are equal and $N\rightarrow\infty$ will $P_{v''=1}=0.5$.} The loss-channel induced by parity mixing becomes negligible for stray electric fields below \SI{1}{\volt\per\centi\meter}. 

\subsection{Two-photon ionisation}
The $A^1\Pi, v'=0$ state of AlF is located 5.45~eV above the $X^1\Sigma^+, v''=0$ state, more than half way up to the ionisation potential at 9.73~eV. Therefore, after excitation on the $A^1\Pi,v'=0\leftarrow X^1\Sigma^+,v''=0$ band near 227.5~nm, a second photon from the same (cw) laser can ionize the molecule, creating an AlF cation in the $^2\Sigma^+$ electronic ground state and a free electron. 

It is quite uncommon to observe such a single color (1+1) resonance enhanced multiphoton ionisation ((1+1)-REMPI) process for molecules using a cw laser. The cross section for the excitation into the ionisation continuum is many orders of magnitude smaller than the cross section of the first, resonant excitation step. The total ionisation yield scales linearly with the time that the molecules spend in the electronically excited state. This makes ionisation with a cw laser difficult to observe, in particular if the intermediate state lives less than 2 ns. However, AlF can be excited on a quasi-cycling transition to increase the time the molecules spend in the excited state to $n_{\text{ph}}\tau$, which can be much larger than $\tau$.

Ionisation leads to a loss from the optical cycle, which can be expressed as
\begin{align}
\gamma_i(I)=\frac{\Gamma_{\textrm{ion}}}{R}=\frac{\sigma_i I\tau}{\hbar\omega},
\end{align}
where $\Gamma_{\textrm{ion}}=\sigma_i \rho_{ee}(I,\delta) I/\hbar\omega$, $\rho_{ee}(I,\delta)$ is the fraction of AlF molecules in the excited state, which depends on the intensity $I$ and the detuning from resonance $\delta$, $\sigma_i$ is the photoionisation cross section from the $A^1\Pi$ state \textcolor{black}{and $R=\rho_{ee}(I,\delta)\Gamma$ is the photon scattering rate}.
The ionisation cross section for the $A^1\Pi$ state is independent of the rotational quantum number $J''$, as is the lifetime $\tau$. Figure \ref{fig:losschannels} b) shows the total number of AlF cations produced as a function of the laser frequency while scanning a high power cw laser over the Q-branch of the $A^1\Pi,v'=0\leftarrow X^1\Sigma^+,v''=0$. The spectrum resembles a power-broadened version of figure \ref{fig:qbranchhfs} a), with a slightly shifted Q(1) peak. The weak electric field mixes the opposite parity states in $A^1\Pi$ which increases the number of accessible levels and therefore the ionisation probability from the $F'=3$ and $F'=4$ states, whose $\Lambda-$doublet components are particularly close. This causes the Q(1) line to appear shifted by about 100 MHz. Moreover, the electric field used to extract the molecular ions from the ionisation region must be low to prevent opening a rotational loss channel (see previous subsection). The ion-yield from (1+1) REMPI decreases with increasing electric field.  

The number of AlF molecules passing through the ionisation volume per pulse is determined by absorption spectroscopy to $N_{\textrm{mol}}=1.2\times 10^8$, with a peak density of $5.5\times 10^7$ cm$^{-3}$. The number of ions detected from the same interaction region is measured using a compact, linear, Wiley-McLaren time-of-flight setup to $N_{\textrm{ion}}=259/\epsilon=650$ per pulse. The ion detection efficiency, $\epsilon\simeq0.4$, is mainly determined by the open area ratio of the microchannel plates \cite{Fehre2018}. The number of detected ions is likely to be underestimated, due to the low electric field (5~V/cm) used to extract the ions.
%move order of magnitude to here
With the laser frequency locked to the center of the Q(1) line, the fraction of the molecules that is ionised by the 17~W/cm$^2$ laser beam (peak intensity) $N_{\textrm{ion}}/N_{\textrm{mol}}=5.4\times10^{-6}$ with $\Gamma_{\textrm{ion}}=N_{\textrm{ion}}/(t_i N_{\textrm{mol}})=1.6$ s$^{-1}$, where $t_i=3.3\times 10^{-6}$ s is the interaction time of the molecules with the laser beam. During this time the molecules scatter on average $\langle n_{\text{ph}}\rangle=60$ photons (see figure \ref{fig:r1-p3-plots} c). With $R=18\times 10^{6}$ s$^{-1}$, the fractional loss from the optical cycle, due to ionisation, becomes $\gamma_i=4.7\times 10^{-9} s$, \textcolor{black}{with $s=I/I_{\textrm{sat}}$ being the saturation parameter,} and $\sigma_i=2\times10^{-18}$~cm$^{2}$. The results presented here should be used as an order-of-magnitude estimate for the loss rate and the ionisation cross section. The low extraction voltage prevents determining the ionisation volume precisely. 

\section{Conclusion}
The AlF molecule holds great prospects for trapping and cooling molecules in a MOT at a density comparable to atomic MOTs. This is an essential step towards using laser-cooled molecules for new studies of strongly interacting, many body systems and precision measurements. We combined a bright molecular beam of a chemically stable molecular species with a new, simple and fast optical cycling scheme that provides an exceptionally high spontaneous scattering force. Both are essential ingredients for a high-density MOT.

By solving the optical Bloch equations numerically for the 72-level system of the Q(1) line of the $A^1\Pi\leftrightarrow X^1\Sigma^+$ transition, we showed that for linear polarisation, the obtainable peak scattering rate is limited to $R=\Gamma/30=17\times10^6$ s$^{-1}$ by the precession rate of coherent dark states. The scattering can be increased further by modulating the polarisation. The numerical results were compared to a rate-equation model which captures the dynamics well for low laser intensities or short interaction times, but fails at high intensities. For low intensities, the rate equations predict a higher scattering rate for higher Q lines because the hyperfine intervals in the $A^1\Pi$ state reduce with increasing $J'$.

We have characterised a bright and slow molecular beam and developed a new method to determine the forward velocity distribution in a single molecular pulse. The number of particles per pulse leaving the source is comparable to atomic buffer gas beam sources, indicating a high production efficiency for AlF. We find that using NF$_3$ as a fluorine donor gas significantly improves the long-term stability of the molecular beam without affecting the beam brightness. The bright molecular beam allows us to deliver about $5\times10^8$ molecules per pulse into the capture volume of the MOT, per shot, per rotational state.

We used this molecular beam to benchmark the simulations against three independent measurements of the optical cycling rate: calibrated laser induced fluorescence, optical pumping into the dark $A^1\Pi, v''=1$ state and deflecting the molecular beam using radiation pressure. The measured scattering rate reaches $R=17(2)\times 10^6$ s$^{-1}$ for the Q(1) line, consistent with the simulations from the optical Bloch equations. Optical pumping and deflection measurements showed that the effective scattering rate increases for higher Q lines. A second UV laser was used to efficiently recover the molecules that were pumped to $v''=1$ and we measured the losses from the optical cycle due to parity mixing and two-photon ionisation. If stray electric fields are well-controlled, approximately $10^4$ photons can be scattered with only two lasers. 

The large spontaneous scattering force measured via molecular beam deflection will result in a very short slowing distance ($\approx 2$ cm) and an exceptionally high capture velocity of the MOT ($\approx 50$ m/s), both essential for a high density MOT of molecules. The electronic structure of AlF allows a straight-forward implementation of a Zeeman slower, which opens the possibility to accumulate molecules from a continuous molecular beam \cite{Shaw2020}. In addition, the spin-forbidden $a^3\Pi\leftrightarrow X^1\Sigma^+$ transition lends itself to narrow-line cooling and precision measurements \cite{Truppe2019}. 

\section{Acknowledgment}
We are grateful to Mike Tarbutt and Micha\l{} Tomza for valuable discussions. We thank the mechanical and electronic work shops of the Fritz Haber Institute for expert technical assistance. This project has received funding from the European Research Council (ERC) under the European Union’s Horizon 2020 research and innovation programme (Grant agreement No. 949119). 
\bibliography{paper}
\end{document}